\documentclass[12pt,preprint]{aastex}

\usepackage{rotating}

\newcommand{\iso}[2]{\hbox{${}^{#1}{\rm #2}$}}
\newcommand{\msun}{\ensuremath{{M}_{\sun}}}

\shorttitle{The $s$-process at low metallicity and CEMP stars}
\shortauthors{Lugaro et al.}

\begin{document}


\title{The $s$ process in asymptotic giant branch stars of low metallicity and the 
composition of carbon-enhanced metal-poor stars}


\author{Maria Lugaro}
\affil{Monash Centre for Astrophysics (MoCA), Monash University,
Clayton VIC 3800, Australia}
\email{Maria.Lugaro@monash.edu}

\author{Amanda I. Karakas}
\affil{Research School of Astronomy \& Astrophysics, Mount Stromlo Observatory,
Weston Creek ACT 2611, Australia}
%\affil{Monash Centre for Astrophysics (MoCA), Monash University,
%Clayton VIC 3800, Australia}
\email{akarakas@mso.anu.edu.au}

\author{Richard J. Stancliffe}
\affil{Monash Centre for Astrophysics (MoCA), Monash University,
Clayton VIC 3800, Australia}
\affil{Research School of Astronomy \& Astrophysics, Mount Stromlo Observatory,
Weston Creek ACT 2611, Australia}
\email{rjs@mso.anu.edu.au}

\and

\author{Carlos Rijs}
\affil{Monash Centre for Astrophysics (MoCA), Monash University,
Clayton VIC 3800, Australia}
\email{cjrij1@student.monash.edu}


\altaffiltext{1}{Monash Centre for Astrophysics (MoCA), Monash University,
Clayton VIC 3800, Australia}


\begin{abstract}

We present models for the $slow$ neutron-capture process ($s$ process) in asymptotic giant
branch (AGB) stars of metallicity [Fe/H]=$-$2.3 and masses 0.9 \msun\ to 6 \msun. We
encountered different regimes of neutron-capture nucleosynthesis increasing in importance
as the stellar mass decreases: the $^{22}$Ne($\alpha$,n)$^{25}$Mg reaction activated
during the thermal pulses, the $^{13}$C($\alpha$,n)$^{16}$O reaction activated in
radiative conditions during the interpulse periods, and the $^{13}$C($\alpha$,n)$^{16}$O
reaction activated during the thermal pulses, also as a result of mild proton ingestion
episodes. The models where the $^{13}$C burns radiatively (masses $\simeq$ 2 \msun)
produce an overall good match to carbon-enhanced metal-poor (CEMP) stars showing
$s$-process enhancements (CEMP-$s$), except they produce too much Na and F. On the other
hand, none of our models can provide a match to the composition of CEMP stars also showing
$rapid$-process enhancements (CEMP-$s/r$). 
The models fail to reproduce the observed Eu abundances, and they also fail to reproduce the correlation 
between the Eu and Ba abundances. They also cannot match the ratio of heavy to light s-process elements 
observed in many CEMP-$s/r$ stars, which can be more than ten times higher than in the solar system. To 
explain the
composition of CEMP-$s/r$ stars we need to invoke the existence of a different ``$s/r$''
neutron-capture process either with features in-between the $s$ and the $r$ processes, or
generated by superpositions of different neutron-capture processes in the same
astrophysical site or in sites linked to each other - for example, in multiple stellar
systems.

\end{abstract}



\section{Introduction} \label{sec:intro}

Following on from the discovery of a large number of halo stars with peculiar chemical 
compositions the last decade has seen much interest sparked in detailed spectroscopic 
observations and nucleosynthetic models of stars of low metallicity 
\citep{beers05,sneden08}. These peculiar compositions provide important constraints on 
the evolution and nucleosynthesis of the low-metallicity stars that existed in 
the early Galaxy. By interpreting the observations in the light of stellar 
nucleosynthesis models we can derive clues as to the sites and nature of the first 
nucleosynthetic processes in the Galaxy. This may ultimately provide important insights 
into the form of the initial mass function of the first stars. 
The composition of metal-poor stars may also provide constraints 
on cold dark matter cosmological models of the formation of galaxies, 
which predict that galactic halos may be composed of different components of possibly  
different composition \citep[e.g.,][]{carollo07}.

Roughly 10-20\% \citep{cohen05,lucatello06,carollo11} 
of old halo stars are enriched in carbon 
([C/Fe]\footnote{The standard spectroscopic notation 
[X/Y]=log$_{10}$(X/Y)$-$log$_{10}$(X/Y)$_{\odot}$ is used 
throughout this paper.}$>$ 1) and are referred to as carbon-enhanced 
metal-poor (CEMP) stars.  Most CEMP stars are also 
enriched in nitrogen, with [N/Fe]$>$1. In most cases [N/Fe]$<$[C/Fe], but also
[N/Fe]$\sim$[C/Fe] and [N/Fe]$>$[C/Fe] are found in roughly 10\% and 5\%, 
respectively, of the total number of CEMP stars.
Roughly 2/3 of CEMP stars also show enrichments in the elements heavier than 
iron, which are produced 
by neutron-capture processes \citep{aoki07,sneden08}.
These enhancements range from pure $s$ enhancements, 
of elements made primarily via $slow$-neutron 
captures (the $s$ process) in the case of ``CEMP-$s$'' stars ([Ba/Fe]$>$1), 
to $s$ and $r$ enhancements, of elements made via the $s$ process and 
the $r$ process, i.e., $rapid$ neutron captures (the $r$ process) in 
the case of ``CEMP-$s/r$'' stars with [Eu/Fe]$>$1 and 
[Ba/Eu]$>$0 \citep{jonsell06}. There are also halo stars with pure $r$ 
enhancements, so-called $r$I (usually classified by setting 
0.5$<$[Eu/Fe]$<$1) and $r$II ([Eu/Fe]$>$1), one 
of which, the famous CS\,22892-052 \citep{sneden03}, belongs to the CEMP 
star family. CEMP stars without enhancements in the elements heavier that 
Fe are commonly referred to as ``CEMP-no'' stars.

To deal with this wide variety of compositions different scenarios for the 
origin of halo stars with peculiar compositions have been proposed. 
Halo stars showing pure $r$ enhancements are believed to 
have been born from material heavily polluted by a primordial core-collapse 
supernova (SN of Type II, SNII) where the $r$-process elements are supposed to 
have formed. Though we note that there are still no SNII models able to reproduce the 
conditions needed for the $r$ process to occur. One of the most favored sites 
to date, the neutrino wind from the nascent neutron star, has recently been 
shown to have major difficulties in producing $r$-process nuclei 
\citep{roberts10}. 
Qualitatively, the C and $s$ enhancements in CEMP-$s$ and CEMP-$s/r$ stars can 
be explained by the presence of an originally more massive binary companion to 
the observed star, which produced C and $s$-process elements during its 
asymptotic giant branch (AGB) phase and transferred them to its companion via 
mass transfer, by either Roche-Lobe overflow or wind accretion 
\citep{lucatello05b}. About 50\% of CEMP stars showing an 
$s$-process signature also show an $r$-process signature and are thus classified as 
CEMP-$s/r$ stars. It is puzzling how CEMP-$s/r$ stars could 
have formed in such a large number given that we believe 
the $s$ and the $r$ processes to be 
independent events occurring in different astrophysical sites, AGB stars and SNII, 
respectively. 
Many scenarios have been proposed \citep[as summarised by][]{jonsell06,lugaro09} 
and most of them seem unlikely. Predictions from a   
scenario where the $r$-process enhancements are 
supposed to be primordial to the binary system have been investigated in detail by 
\citet{bisterzo11a} and will be discussed in \S~\ref{sec:baeu}.

The N enhancements in CEMP stars are also mysterious. The highest enhancements, [N/C]$>$0, 
may be due to the effect of proton captures at the base of the convective envelope
in AGB stars with masses higher than $\simeq$ 3 \msun\ 
\citep[hot bottom burning, HBB][]{lattanzio91,boothroyd93}, which very 
efficiently convert C into N. The milder enhancements can be qualitatively explained by H 
and He burning occurring together when protons are ingested in the He-burning convective 
regions of AGB stars \citep{campbell08,iwamoto04}, or by the operation of some form of deep 
mixing of the envelope material of the AGB star companion down to regions where H burning 
occurs. No physical mechanism for this deep mixing 
has been shown to be capable of producing the required nitrogen enhancement. 
\citet{stancliffe10} 
has shown that it cannot be caused by thermohaline mixing.
Another issue that has received much attention is the question of whether 
material transferred from the AGB companion onto the star now observed as a CEMP star was 
diluted on the secondary star via thermohaline mixing (or any non-convective 
process), or kept at the surface until the star reached the giant branch and developed an 
extended convective envelope \citep{stancliffe07a,stancliffeglebbeek08}.

In this context, it is obvious that detailed models of the $s$ process in AGB 
stars of low metallicity are crucial to improving our understanding of CEMP stars. 
Some models have been presented so far by \citet{vaneck03}, 
\citet{cristallo09a}, and \citet{bisterzo10}
and have been compared to stellar  
observations by \citet{bisterzo11a,bisterzo11b}. With this paper we greatly extend the 
range of published models of the $s$ process in AGB stars by 
presenting the evolution and nucleosynthesis in a large set of stellar  
models of masses ranging from 0.9 \msun\ to 6 \msun\ 
at the typical metallicity of CEMP stars [Fe/H]=$-$2.3. 
The $s$-process predictions for 
low-metallicity AGB stars of masses 
lower than 1.3 \msun\ and higher than 2 \msun\ are provided here for the first time in the literature.
To test different physical and numerical approaches we 
computed stellar evolutionary sequences using two different 
evolutionary codes. We then input these different evolutionary sequences 
into the same post-processing code, which employs a nuclear network 
from H to Bi and 
calculates nucleosynthesis by coupling mixing and burning in convective regions. We then 
compare our model predictions to 
the abundances of CEMP stars as a population using the recent 
compilation of CEMP star compositions from \citet{masseron10}. 

\subsection{The $s$ process in AGB stars of low metallicity}
\label{sprocessintro}

From centre to surface an AGB star consists of: a C-O degenerate core, a He-burning 
shell, a He-rich intershell, a H-burning shell, and an extended convective 
envelope, which is constantly eroded by strong stellar winds. When the whole 
envelope is lost the hot stellar core is left to evolve into the post-AGB phase, 
then it becomes the central star of a planetary nebula and eventually a cooling 
white dwarf.
For a detailed review on AGB evolution we refer the reader to \citet{herwig05}.
A peculiarity of the AGB evolution is that the H- and the He-burning shells are 
activated alternately. Most of the time the H-burning shell is active and 
freshly made He increases the mass of the intershell for periods of time of around 
10$^3$-10$^5$ yr, depending on the stellar mass, with shorter periods
associated with more massive AGB stars.
The increasing mass of the intershell leads to an increase in the temperature 
at its base, until He burning is suddenly triggered resulting in a 
{\it thermal pulse} (TP). 
The energy released in this first phase of He-shell burning cannot be carried out 
radiatively and a convective region develops in the intershell, which lasts a few 
hundreds years (the exact duration depends on mass) and normally extends up to
just below the H-burning shell. 
In low-mass and low-metallicity AGB stars this convective region 
can extend into the H-rich layer and peculiar nucleosynthesis
is expected to occur there \citep{campbell08}, including $s$-process 
nucleosynthesis \citep{cristallo09b}. 
Mild proton-ingestion events of this type are found in our low-mass models. They 
have a strong impact on the $s$-process distributions and will be discussed 
in detail in \S~\ref{sec:results}. 

A TP causes the star to expand, which in turn causes the H-burning shell to be 
quenched. At this stage, the convective envelope can penetrate into the underlying layers and carry 
to the stellar surface material processed by H and He burning (a process referred to as the {\it third 
dredge-up}, or TDU for short), while He burning carries on radiatively. Eventually, He burning is also 
quenched and the whole cycle of H burning, TP, and TDU starts again. This cycle is repeated a few to 
hundreds of times depending on the time it takes for the mass loss to erode the envelope, which in 
turn depends on the initial stellar mass, and possibly the metallicity. The best evidence that TPs 
and the TDU occur in AGB stars is the formation of carbon stars, which show C$>$O at their surface as 
well as the presence of the radioactive element Tc \citep{wallerstein98}. This observed composition 
is explained by the combined effect of partial He burning during TPs (which produces more C than O) 
and the $s$ process (which produces Tc), along with the TDU, which carries a fraction of the 
intershell to 
the stellar envelope. As mentioned above, HBB is present in AGB stars with masses higher than 
$\simeq$ 3 \msun\ at the metallicity considered here, converting C into N.

The $s$ process occurs in the intershell of AGB stars where He is abundant and 
($\alpha$,n) reactions can be efficiently activated. 
For a detailed review on the $s$ process in AGB stars we refer the reader to \citet{busso99}.
There are two possible 
neutron sources: the \iso{13}C($\alpha$,n)\iso{16}O and the 
\iso{22}Ne($\alpha$,n)\iso{25}Mg reactions. The \iso{22}Ne neutron source is 
activated at temperatures higher than roughly 300$\times 10^{6}$\,K (hereafter MK).
These high temperatures can only be reached inside the convective TPs, and 
mainly in AGB stars more massive than $\simeq$ 3 \msun\ \citep{iben77}. 
The \iso{22}Ne is produced via double $\alpha$ captures on \iso{14}N when 
the H-burning ashes are engulfed in the TP. The resulting neutron flux 
occurs over a timescale of the order of a few years with relatively high 
neutron densities, up to 10$^{14}$ n/cm$^{3}$ \citep{gallino98,lugaro03b,vanraai11}. 
The \iso{13}C neutron source is instead activated at temperatures as low as 
90MK \citep{cameron55} but there is not enough \iso{13}C in the H-burning 
ashes of AGB stars to make it an efficient neutron source. Without this 
low-temperature neutron source it is not possible to explain the 
$s$-process enhancements which have been observed in AGB stars since the 1950s 
\citep{merrill52,smith86b}. This is because most of the observed 
$s$-process-enriched AGB stars are of relatively low mass ($<$ 4 \msun) 
where the temperature is not high enough to activate the \iso{22}Ne neutron 
source. Thus, it has become standard practice to artificially include the \iso{13}C 
neutron source in AGB models of the $s$ process. 

The most likely way to form the required \iso{13}C is to allow some extra mixing of protons to occur 
following each TDU episode at the deepest extent of the convective envelope, where a sharp composition 
discontinuity forms at the interface between the H-rich envelope and the He-rich intershell. This 
extra mixing may be physically explained by the effect of semi-convection \citep{hollowell88}, 
hydrodynamical overshoot \citep{herwig00,cristallo09a}, or gravity waves \citep{denissenkov03a}. 
However, there is no agreement yet on which of these proposed mechanisms is responsible for the 
extra mixing and thus we do not know what its detailed features are. The protons are captured by the abundant \iso{12}C, 
producing a layer rich in \iso{13}C and \iso{14}N 
\citep[the $^{13}$C $pocket$,][]{hollowell88,gallino98,goriely00,lugaro03b,cristallo09a}. 
The resulting \iso{13}C usually burns under 
radiative conditions via \iso{13}C($\alpha$,n)\iso{16}O before the onset of the following TP 
\citep{straniero95,gallino98}. This burning releases a large number of free neutrons on a long 
timescale (around $10^4$ yr) and with low neutron densities $< 10^8$ n/cm$^3$. In the region of the 
pocket where \iso{14}N is more abundant than \iso{13}C, no $s$-process nucleosynthesis occurs owing 
to the dominance of the \iso{14}N(n,p)\iso{14}C reaction over neutron captures by Fe seed nuclei and 
their progeny.  In the intermediate-mass AGB stars that experience HBB, the formation of a \iso{13}C 
pocket may be inhibited by proton captures occurring at the hot base of the convective envelope 
during the TDU, 
which produce \iso{14}N and not \iso{13}C \citep{goriely04}. Extremely deep TDU may also 
inhibit the activation of the \iso{13}C($\alpha$,n)\iso{16}O reaction by penetrating into 
regions of the stellar core with a low abundance of \iso{4}He \citep{herwig04a}. While these 
processes do need further investigation, on first sight a dichotomy appears in models of the $s$ 
process in AGB stars where the \iso{13}C neutron source is activated in the low-mass models, while 
the \iso{22}Ne neutron source is activated in the higher-mass models. At the metallicity we consider 
here the mass at which the importance of the two neutron sources switches is $\simeq$ 3 \msun\ 
\citep{goriely04}.

An interesting feature of the \iso{13}C neutron source is that it is a 
primary neutron source because it is produced from the H and He 
originally present in the stars. Since the 
total time-integrated neutron flux $\tau$ from the 
\iso{13}C source is proportional to \iso{13}C/$Z$, 
the $s$-process predictions have a strong dependence on the metallicity of the star
\citep{clayton88,busso01}. Specifically, with $\tau$ increasing when decreasing $Z$, 
heavier elements are produced at lower metallicities. This property allowed 
\citet{gallino98} to predict the existence of low-metallicity Pb-rich stars, which was 
confirmed by observations of Pb-rich CEMP stars \citep[e.g.,][]{vaneck03}. 
On the other hand, the \iso{22}Ne neutron source is traditionally considered 
to be a secondary neutron source, since it relies on the original CNO 
nuclei present in the star. However, at the low metallicities 
considered here \iso{22}Ne also has an important primary component because 
the N abundance in the H-burning ashes increases significantly from its 
initial value. This is due to the effect of TDU carrying primary \iso{12}C to the 
envelope, which is then converted into N by H burning. 

The \iso{13}C and the \iso{22}Ne neutron sources produce neutron fluxes and 
neutron-capture nucleosynthesis with different features. In summary, it 
has been shown that the \iso{13}C source is responsible for the 
production of the bulk of the $s$-process elements in low-mass AGB stars
reaching, as mentioned above, up to Pb at low metallicities \citep{gallino98}. 
Branching points on the $s$-process path are mostly closed during this neutron 
flux. The \iso{22}Ne source instead produces smaller abundances of
the $s$-process elements but it efficiently activates branching points along 
the $s$-process path \citep{abia01,vanraai11}. Note that at low metallicity 
\iso{22}Ne is both a neutron source and a neutron poison via (n,$\gamma$) 
reactions, leading to the production of light elements such as Na and Mg 
\citep{herwig04b,karakas10a,bisterzo10}. Based on our models in \S~\ref{sec:results} we 
revise the different neutron-capture scenarios for AGB stars of low 
metallicity.

\section{Stellar Models} \label{sec:models}

The stellar evolutionary sequences were computed using the {\sc stars}  
and the Monash/Mount Stromlo (hereafter Stromlo) codes. 
The {\sc stars} code was originally written by \citet{eggleton71} and has been 
updated by many authors \citep[e.g.][]{pols95,stancliffeeldridge09}. Mixing is 
treated via a diffusion equation \citep{eggleton72} and convective boundaries are determined via the 
Schwarzschild criterion. The opacities tables are 
from \citet{eldridgetout04} and the nuclear reaction rates are 
from \citet{stancliffe05b}. Models are evolved from the pre-main 
sequence to the AGB using 999 mesh points. Mass loss before the AGB phase is 
included using the 
\citet{reimers75} prescription, with $\eta=0.4$. Mass loss during the AGB phase is 
included using the prescription of \citet{vw93}. 
The computation of the AGB phase with the {\sc stars} code has been described in detail in
\citet{stancliffe04b}. We included an approximate
contribution to the molecular opacities \citep[based on the work of][] 
{marigo02} that varies with envelope composition 
\citep{stancliffeglebbeek08}. Convective overshooting is not included at any stage 
in the evolution, and a mixing length of $\alpha=2.0$ is employed throughout.

The Stromlo code uses the standard mixing-length theory for convective
regions, with a mixing-length parameter $\alpha =1.75$, and
determines the border of convective regions by applying the Schwarzschild criterion. 
Convective overshooting is not employed at any stage 
in the evolution, although the code searches for a neutral border to 
the convective zone in the same way as described in \citet{lattanzio86}. 
This can have the effect of increasing the
efficiency of the TDU \citep{frost96,karakas02}.  At high temperatures the 
OPAL opacities are used \citep{iglesias96}, and at low temperatures 
(below $T \le 10^{4}$K) two different prescriptions are employed.
The models between 1 to 2.5$M_{\odot}$ \citep{karakas07b}
include an approximate treatment for the molecular opacities
(in particular CN, CO, H$_{2}$O, TiO) using the formulations from
\citet{bessell89} and corrected by \citet{chiosi93}. These fits do
include some compositional dependence, but do not account for large 
variations in the envelope C/O ratio or N abundance. 
In the model of 0.9 \msun\ and models between 3 to 6 \msun\ 
the code uses the (scaled-solar) low-temperature opacity tables 
from \citet{ferguson05} in place of the \citet{bessell89} fits.
Mass loss before the AGB phase is included using the 
\citet{reimers75} prescription, with $\eta=0.4$. For the models
between 0.9 to 2.5 \msun\ mass loss during the AGB phase is 
included using the prescription of \citet{vw93}. For models of
3 to 6 \msun\ mass loss during the AGB is included using the
\citet{reimers75} prescription with $\eta$ values that vary with
mass \citep[see discussion in][]{karakas10a}. 

Two further Stromlo models of 2 \msun\ and 2.5 \msun\ of $Z=0.0001$ were
calculated using the low-temperature C and N-rich opacities from \citet{lederer09}.
These new stellar evolutionary sequences were computed with \citet{vw93}
mass loss on the AGB and updated reaction rates for the
\iso{14}N($p,\gamma$)\iso{15}O reaction
\citep[the LUNA rate from][]{bemmerer06}, and the NACRE rate for
the triple-$\alpha$ process \citep{angulo99}.
These models were computed to quantitatively test the effect of the low-temperature
opacity tables on the evolution of low-metallicity AGB stars, in particular
to reconcile the number of TPs obtained between the Stromlo
and the {\sc stars} models. 
Many of the important features of the new
models are the same as the old models, including the maximum depth of the
TDU, however, the main change is a reduction in the number of TPs
and hence the total amount of He-shell material dredged to the envelope over the
AGB phase. One interesting feature of the new models is that they
were both evolved to the white dwarf cooling track, where the final
envelope masses are a few $\times 10^{-5}$ \msun, providing an excellent
estimate of the total number of predicted TPs.

All our models were computed using a metallicity of [Fe/H]=$-$2.3. The peak 
of the distribution of the metallicities of CEMP stars enhanced in 
neutron-capture elements is at [Fe/H]=$-$2.3 and the overall range cover
$-3.5<$[Fe/H]$<-$1. \citep[see, e.g., Fig. 7 of][]{aoki07}. Keeping this caveat in 
mind, we will compare our models to the composition of CEMP stars as a 
population, rather than to single objects. 

\begin{figure} 
\begin{center}
%\epsscale{.80} 
%\plotone{BaEu_Full.ps} 
\includegraphics[scale=0.6,angle=0]{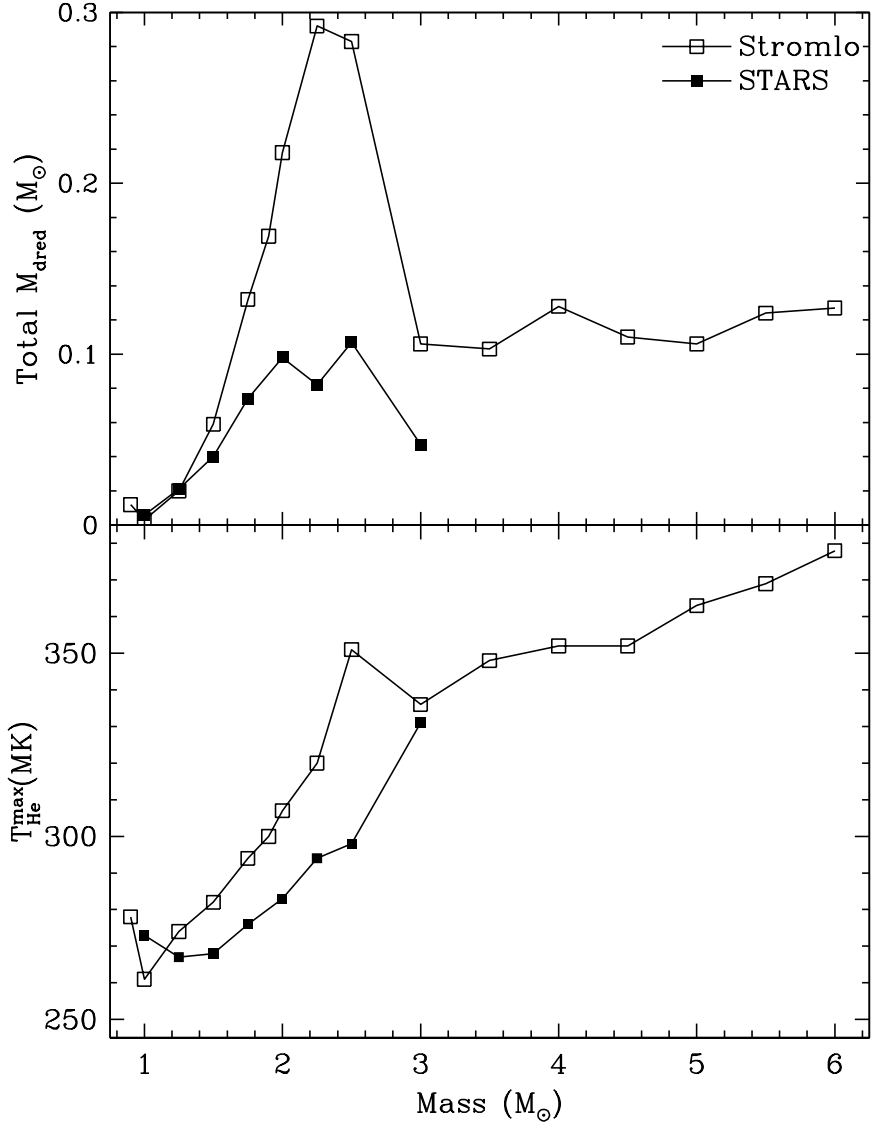} 
\caption{The total mass dredged-up by the TDU 
(top panel) and the maximum temperature at the base of TPs (bottom panel) as a function 
of the initial stellar mass for our AGB models. The change in the trend for the Stromlo models
between 2.5 \msun\ and 3 \msun\ is caused by the change in the adopted
mass-loss law (see the text for details). The use of the
\citet{reimers75} mass-loss prescription for stars with masses 3
\msun\ -- 6 \msun\ has the effect of shortening the AGB lifetime and
reducing the number of TDU episodes.}
\label{fig:stellardata}
\end{center}
\end{figure}

The main features of our stellar models are presented in Fig.~\ref{fig:stellardata}, and 
Tables~\ref{tab:starscode} 
and \ref{tab:stromlocode}. The tables include: the initial stellar mass, the total 
number of thermal pulses (TPs), which thermal pulses are followed by the TDU (TPs 
with TDU), the maximum temperature in the TPs followed by the TDU (T$^{\rm max}_{\rm He}$,
also plotted in the bottom panel of Fig.~\ref{fig:stellardata}),
the maximum TDU efficiency $\lambda_{\rm max}$, defined as the ratio of the 
mass scooped up to the envelope by the TDU over the mass growth of the 
H-exhausted core in the previous interpulse phase, 
the total mass dredged-up by TDU during the TP-AGB (total M$_{\rm dred}$,
also plotted in the upper panel of Fig.~\ref{fig:stellardata}), the mass of the envelope at the 
end of the computed evolution (final M$_{\rm env}$), if HBB is at work or not, and the 
maximum temperature at the base of the convective envelope.
The main features arising from the analysis of the two 
sets of models are:

\begin{enumerate}

\item{The number of TPs (Column 2) is smaller - roughly half at any given initial 
mass (Column 1) - in the {\sc stars} code than in the Stromlo code. This is 
because the {\sc stars} code includes an approximation to the molecular opacities that 
varies with the composition of the envelope. This is known to 
affect the mass-loss rates and shorten the stellar lifetime \citep{marigo02}. 
The Stromlo 2 \msun\ and 2.5 \msun\ models computed using 
low-temperature opacities with varied C and N abundances from \citet{lederer09} 
experienced $\simeq$ 40\% fewer TPs thus confirming the effect of shortening 
the stellar lifetime when more realistic low-temperature 
opacities are used. The 2 \msun\ model of the same metallicity presented by 
\citet{cristallo09a} uses the same molecular opacities from \citet{lederer09}, but a slightly 
different mass-loss law and 
has 15 TPs, close to the number of TPs in our 2 \msun\ models.}

\item{While the overall trend is for the number of TPs to increase with initial 
stellar mass (as a larger envelope means that it takes more 
time for the mass loss to erode it), models with masses lower than 1.25 \msun\ show 
instead an increase in the number of TPs. We obtain a higher total TDU 
mass and a higher intershell temperature in the 0.9 \msun\ model than in the 1 \msun\ models. 
This result is most likely to be affected by uncertainties in the mass-loss rate, 
which are large, and needs to be investigated in more detail. In the lowest-mass models 
the initial pulses are too weak to trigger efficient TDU (Column 3), and thus C/O$>$1 is 
reached later during the evolution.}

\item{The fact that the {\sc stars} models experience fewer TPs with TDU (Column 3) than the
Stromlo models leads to a smaller amount of dredged up material being mixed into the
envelope over the AGB phase (Columns 6). This is illustrated in Fig.~\ref{fig:stellardata}.
The exception to this is the 1 \msun\ model. The total TDU mass does
not change linearly with the number of 
TPs because it is also determined by the numerical treatment of the inner border 
of the convective envelope, which differs in the two codes. 
More efficient TDU is typically favored in the {\sc stars} models (Column 5).}

\item{The maximum temperature reached in TPs followed by the TDU (Column 4) is 
lower in the {\sc stars} models than in the Stromlo models (except for the {\sc stars} 
1 \msun\ model, which has more TDU episodes). 
The \iso{22}Ne($\alpha$,n)\iso{25}Mg reaction starts to operate at temperatures
of $\simeq$300 MK, which is reached in the 1.9 \msun\ Stromlo model and in the 2.5 
\msun\ {\sc stars} model.
This quantity is also known to increase with the TP 
number.}

\item{The final envelope mass is mostly determined by how long we managed to 
run the codes before encountering insurmountable convergence problems, usually during 
the final phase of very high mass loss that is typical of the prescription of 
\citet{vw93} we used. It is usually easier to get closer to the end 
of the AGB phase with models of lower masses, although the
intermediate-mass evolutionary sequences computed by the Stromlo code 
were also evolved to small (around $0.1$ \msun) envelope masses probably due 
to the choice of the \citet{reimers75} mass loss
prescription on the AGB, instead of \citet{vw93}. For these models
we do not expect any further TPs and TDU
episodes, except perhaps at the very tip of the AGB or during 
the post-AGB phase.}

\item{The minimum mass at which we see the operation of HBB is 3 \msun, the same for 
both codes. HBB is very mild in this Stromlo model, with a maximum temperature 
at the base of the convective envelope $\simeq$40 MK (Column 8), while it is more 
efficient for the {\sc stars} 
(T$^\mathrm{max}_\mathrm{bce}\simeq$58 MK). HBB is already very efficient  
for the 3.5 \msun\ Stromlo model, with a maximum temperature $\simeq$76 MK.}

% AK - This last item is a repeat of items 3 and 4
% and should be merged with them. 

\item{The Stromlo models with mass $>$ 3 \msun\ reach temperatures in the He-burning shell 
up to 378 MK. These temperatures are high enough for an efficient activation of the 
$^{22}$Ne neutron source. The total TDU mass dredged up is similar in all intermediate-mass 
models, with a slight trend to increase with increasing stellar mass due to the increasing 
number of TPs.} 

\end{enumerate}

We note that most of the quantities presented in Tables~\ref{tab:starscode} and 
\ref{tab:stromlocode} are affected by uncertainties related to the treatment of the TDU and to 
the mass-loss rate.

\subsection{Nucleosynthesis models}
\label{sec:nucmodels}

To study the nucleosynthesis of the elements up to Bi we have used a 
post-processing code that takes as input the stellar evolutionary sequences 
described above. The stellar inputs are the temperature, density, and convective 
velocities (for convective regions) at each mass shell in the star as a function 
of time during the life of the star. Convective velocities are needed because the 
code includes both the changes due to nuclear reactions and those due to mixing 
in the equations to be solved for the abundances. An implicit method is used to 
solve N equations, N being the number of species included in the nuclear network, 
by inverting a large matrix containing N$^2$ elements. The code has been 
previously described in detail by, e.g., \citet{cannon93}, \citet{karakas07b}, 
and \citet{lugaro04}.

Given that the number of species N determines how large the matrix to be 
solved is, in the interest of making the computation relatively fast (from a few 
days to a few weeks, depending on the initial stellar mass), we used for this 
work a basic $s$-process network, where we included a total of 320 species 
mostly along the valley of $\beta$ stability, and 2,336 reactions. We based our 
nuclear reaction network on the JINA REACLIB database, as of May 2009. From 
this database, we updated the rates for the neutron source reactions: we took 
the \iso{13}C($\alpha$,n)\iso{16}O rate from \cite{heil08} and the 
\iso{22}Ne($\alpha$,n)\iso{25}Mg from \cite{karakas06a}.

For the initial composition of most models we used the solar distribution of 
abundances from \citet{asplund09} scaled down to [Fe/H]=$-$2.3. 
Solar abundances of C, N, O, Ne, Mg, Si, S, Ar, and Fe are the 
pre-solar nebula values from Table~5 of \citet{asplund09}; F is 
the meteoritic value of $\log \epsilon$ (F)$_{\odot}$ = 4.42 
from Table~1 of the same paper (chosen because it has a lower uncertainty), 
and for many of the elements heavier than Fe we use the meteoritic values 
for the solar abundances (e.g., Sr, Eu, Pb).

We also computed some models by varying the initial composition to: (1) the values 
predicted by galactic chemical evolution models for solar neighborhood stars 
of metallicity [Fe/H]=$-$2.3 from \citet{kobayashi11}, this case affects only the initial 
composition of the elements up to Zn; (2) the values obtained assuming that the 
star was born with an $r$-process enrichment and in some cases also an $s$-process 
enhancement. Enhancements in the initial $r$-process abundances may have arisen 
from the pollution from one or a few primordial SNeII. Enhancements in the 
initial $s$-process abundances may have arisen from pollution from primordial 
massive stars producing $s$-process elements during core He and shell C 
burning \citep{pignatari08}, 
or via the still unidentified ``light element primary process'' 
\citep{travaglio04,montes07}. To compute these $r$ and $s$-process 
enhancements we used the $s$ and $r$ contributions given in \citet{sneden08}.

\subsection{The inclusion of the \iso{13}C pocket}
\label{sec:pocket}

The methodology for the inclusion of the \iso{13}C pocket deserves specific attention as 
the formation of this neutron source is still the main uncertainty in the $s$-process 
models. For reasons discussed above (\S~\ref{sprocessintro}), we introduced a 
\iso{13}C pocket for models of masses $<$ 4 \msun, while we did not introduce it for the 
higher masses, except in one test case, where we introduced a \iso{13}C pocket in the 
Stromlo 5.5 \msun\ model. 
Also the 3 \msun\ model was computed with and without the inclusion of the \iso{13}C 
neutron source to illustrate in detail the different effects of the two neutron sources.
This mass lies where the two regimes switch: the $s$-process dominated by the \iso{13}C 
neutron source and the $s$-process dominated by the \iso{22}Ne neutron source.

The inclusion of the \iso{13}C neutron source was performed artificially during the 
post-processing by forcing the code to mix a small amount of protons from the 
envelope into the intershell at the deepest extent of each TDU
episode. All the studies to date 
investigating the formation of the \iso{13}C pocket by various mechanisms have found 
that the proton abundance in the intershell decreases monotonically. Hence, we apply 
the basic assumption that the proton abundance in the intershell decreases 
monotonically from the envelope value of $\simeq$ 0.7 to a minimum value 
of 10$^{-4}$ at a given point 
in mass located at ``M$_{\rm mix}$'' below the base of the envelope. This method is 
described in some more detail in \citet{lugaro04} and is very similar to that used by 
\citet{goriely00}. We made the further choice that the proton abundance 
decreases exponentially, i.e., linearly in a logarithmic scale. Studies of the 
formation of the \iso{13}C pocket have found profiles that can slightly differ from 
this basic assumption, as well as from each other. For example, compare the profile shown
in Fig.~4 of \citet{herwig00} to the profiles shown in Figs.~1-4 of 
\citet{cristallo09a}. \citet{goriely00} tested the effect of using 
a ``slow'' or a ``fast'' profile (see their Fig.~10) and found no major differences 
in the $s$-process distribution. 

We tested different values of our parameter M$_{\rm mix}$, 
% AK - this is not correct. Mmix is the extent in mass of the proton
% profile, the mass of the c13 pocket < Mmix
%
% representing the extent in mass of the \iso{13}C pocket. 
representing the extent in mass of the proton profile. The extent in
mass of the resulting \iso{13}C pocket is $<$M$_{\rm mix}$ 
\citep{goriely00,lugaro03b,cristallo09b}.
For some models of mass $<$ 3 \msun\ we tested 
M$_{\rm mix}$=0, 0.0006 \msun, 0.001 \msun, 0.002 \msun, and 0.004 \msun. For the 3 
\msun\ models we used M$_{\rm mix}$= 0, 0.0005 \msun\ and 0.001 \msun. For the 5.5 
\msun\ model, we tested M$_{\rm mix}$=0 and 0.0001 \msun. To keep the standard 
assumption that the \iso{13}C pocket should be a minor fraction ($\simeq$1/10 - 
1/20) of the intershell mass \citep{gallino98,goriely00} we had to decrease the value of 
M$_{\rm mix}$ with the stellar mass because the mass of the intershell is at least
an order of magnitude smaller in the intermediate-mass AGB stars than in the 
lower-mass models: around $10^{-2}$ \msun\ in a 1.5 \msun\ model 
compared to around $10^{-3}$ \msun\ in a 6 \msun\ model. 

\section{Results} \label{sec:results}

\begin{figure}
 \begin{center}
%\epsscale{.80} 
%\plotone{BaEu_Full.ps} 
\includegraphics[scale=0.6,angle=0]{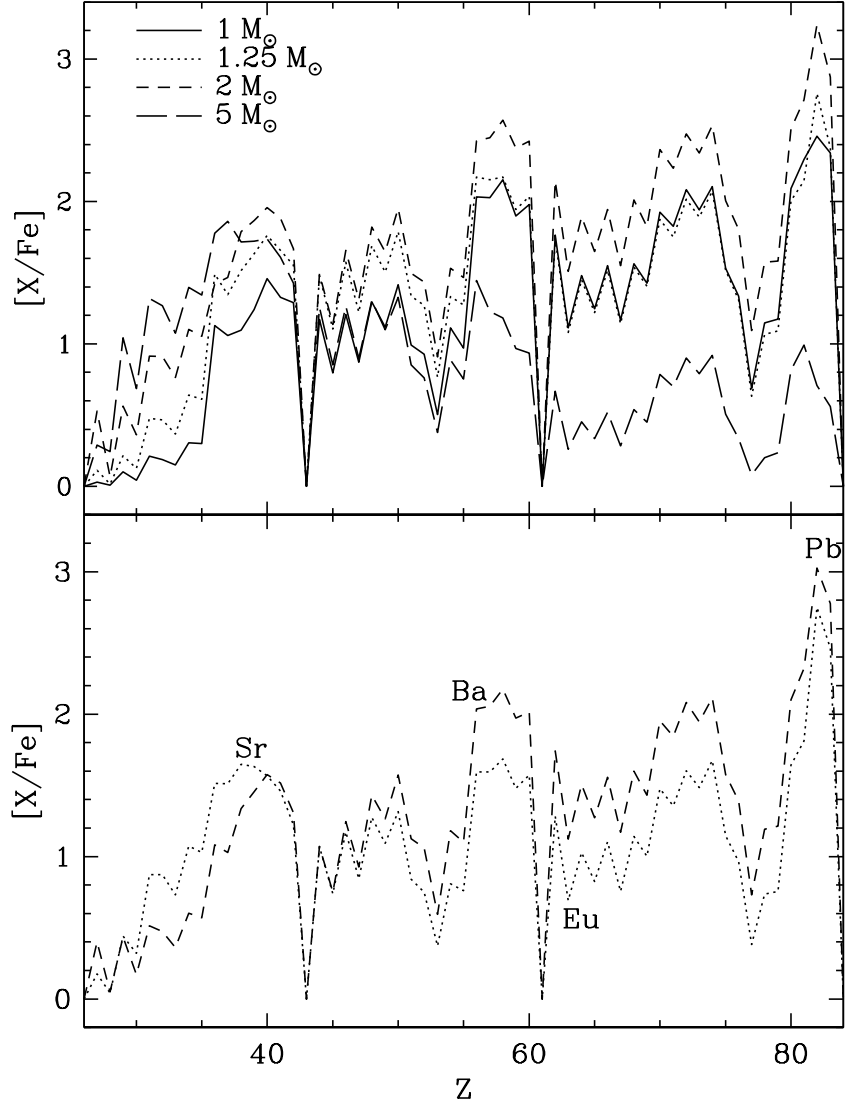} 
\caption{The [X/Fe] ratio as function of the atomic number Z 
for elements heavier than Fe for selected Stromlo (upper panel) 
and {\sc stars} (lower panel). Data are surface abundances after 
the last computed TP. The selected models 
represent cases where the neutron-capture nucleosynthesis occurs 
according to the four regimes: The 5 \msun\ model represents the case where 
all neutrons are released by the $^{22}$Ne neutron source in the TPs (Regime 1).
The 2 \msun\ models represent the regime where most
neutrons are released under radiative conditions during the interpulse periods
by the $^{13}$C neutron source (Regime 2). 
The {\sc stars} 1.25 \msun\ model 
represent a case where neutrons  released in radiative conditions by
the $^{13}$C neutron source primarily determine [Pb/Ba], while [Ba/Sr]
is also affected by $^{13}$C burning convectively (Regimes 2, 3, and 4). 
The Stromlo 1.0 \msun\ and 1.25 \msun\ models 
represent the regime where most of the neutrons are released in the convective TPs 
by the $^{13}$C neutron source (Regimes 3 and 4). 
All the other models show behaviour in between the selected models 
and are not included for the sake of clarity 
(all the data can be found in the on-line tables).}
\label{fig:heavyresults}
\end{center}
\end{figure}

The results in terms of [X/Fe] for selected key elements (where X represent a given 
element) for all our models are presented in Tables~\ref{tab:resultsSTARS} and 
\ref{tab:resultsMM} and Figs.~\ref{fig:heavyresults} and 
\ref{fig:lightresults}\footnote{We had numerical problems computing the 
nucleosynthesis for the {\sc stars} 1 \msun\ model, which we still have not manage to 
solve, so the results of this model are not included in the 
tables and figures.}. The elemental 
abundances [X/Fe] in these tables are given at the stellar surface at the end of 
the computed evolution. The extended tables available 
on-line list, for each element X included in the network: log 
$\epsilon$(X), i.e., 12 + log$_{10}$ \{N(X)/N(H)\}, [X/H], [X/Fe], and
[X/O], and X(i), which 
is the fraction of element i by mass at the surface. We also include the
C$/$O ratio and the isotopic ratios of C, N, O, and Mg. All elemental
and isotopic ratios are number ratios. These quantities are tabulated at 
the stellar surface at the start of the evolution, after each TP, and at the end of the 
evolution, accounting for the radioactive decay of long-lived isotopes.

We also present in the extended tables and in the figures of 
\S~\ref{sec:compare} the $s$-process indices light ``ls'' and heavy ``hs''. We 
choose the three main elements belonging to the first $s$-process peak Sr, Y, and 
Zr to define [ls/Fe]=([Sr/Fe]+[Y/Fe]+[Zr/Fe])/3 and the three main elements 
belonging to the second $s$-process peak Ba, La, and Ce to define 
[hs/Fe]=([Ba/Fe]+[La/Fe]+[Ce/Fe])/3. We choose these definitions to
compare our models to the data from the compilation of \citet{masseron10}, who 
report abundances for Ba, La, Ce, Eu, and Pb\footnote{For data on the ls 
elements we searched the SAGA database \citep{suda08}.}. We note that 
our definitions are different from those used by \citet{bisterzo10} who took Y and Zr 
as the ls elements, and La, Nd, and Sm, as the hs elements. That choice 
was made in order to select the elements for which spectroscopic determinations 
are more reliable\footnote{Note that, according to \citet{arlandini99}, about 80\% of the solar 
abundances of Sr, Y, Zr, Ba, and Ce are produced by the $s$ process, while La, Nd, 
and Sm have an important $r$-process contribution: 62\%, 56\%, and 29\% of their solar
abundances are ascribed to the $s$ process, respectively.}. 
In the extended on-line tables we provide 
[ls/Fe] and [hs/Fe] as defined in this work and as defined by \citet{bisterzo10}, 
for comparison.

We have identified four different regimes of neutron captures in our 
models, which explain the different abundance distributions for the 
elements heavier than Fe in Fig.~\ref{fig:heavyresults}, corresponding to the  
different [Ba/Sr] and [Pb/Ba] ratios listed in 
Table~\ref{tab:resultsSTARS} and \ref{tab:resultsMM}.
These regimes are described 
below, together with the typical stellar mass ranges at which they occur. 
The mass range in which each regime is at work is slightly different in the two sets of 
models, {\sc stars} and Stromlo and 
in most models more than one regime is at work. In principle   
the final results depend on the interplay of the different regimes, though  
often a dominant situation can be identified. The regimes are:

\begin{enumerate}

\item{{\it The $^{22}$Ne($\alpha$,n)$^{25}$Mg neutron source operates inside TPs 
when the temperature at the base of the TP reaches above 300 MK.} From 
Fig.~\ref{fig:stellardata} it is clear that this neutron 
source operates when the initial stellar mass is higher than about 2 -- 2.5 
\msun. Two key features of this regime are that (i) the temperature is never high 
enough to completely burn the $^{22}$Ne nuclei (at most the $^{22}$Ne 
intershell abundance 
decreases by a factor of three) and (ii) the neutrons are distributed 
over the whole intershell by the convective TP. The resulting overall number of 
neutrons per Fe seed produces distributions where lighter $s$-process elements 
are favored with respect to heavier $s$-process elements. For
this reason the [Ba/Sr] and [Pb/Ba] ratios are negative. 
The neutron density reaches up to $\simeq 10^{14}$ 
n/cm$^{3}$, so branching points are affected but the overall distribution 
is a typical $s$-process distribution. In our models the 
activation of the $^{22}$Ne neutron source in intermediate-mass AGB stars does not lead 
to an enhanced production of Eu, as was suggested by \citet{masseron10}. The 
[Ba/Eu] ratio in these models 
remains close to unity\footnote{The [Ba/Eu] ratio decreases down to 0.11 
only in the 3 \msun\ {\sc stars} 
model computed without the inclusion of the $^{13}$C pocket, but this is due to the 
fact that not much $s$ process occurs here so that very little Ba and no Eu are 
produced in this model.}. 
This is the value expected by the $s$ process in equilibrium with 
$\sigma_{\rm A} N_{\rm A} = constant$, where $\sigma_A$ is the neutron-capture
cross section of the isotope of mass A and $N_{\rm A}$ is its $s$-process abundance, 
considering that the neutron-capture cross section of, e.g., the 
neutron magic nucleus $^{138}$Ba is $\sim$650 times lower than 
the neutron-capture cross section of, e.g., $^{153}$Eu, while its solar abundance 
is $\sim$65 times higher\footnote{We can approximately apply the $\sigma_{\rm A} N_{\rm A} 
= constant$ rule because the Eu abundance 
in $s$-process conditions follows that of Ba, the magic nuclei that precedes it in mass.}.}

\item{{\it The $^{13}$C($\alpha$,n)$^{16}$O neutron source operates in 
radiative conditions during the interpulse periods, with $^{13}$C produced by 
the inclusion of a $^{13}$C pocket (cases with M$_{\rm mix} \neq$ 0).} Typical 
models where this is the main regime of neutron captures are the models with 
initial masses between roughly 1.75 \msun\ and 3 \msun. Two key features of this 
regime are that (i) the $^{13}$C nuclei burn completely in radiative conditions 
and (ii) the neutrons are captured locally in the thin $^{13}$C-pocket layer. 
The overall number of neutrons per Fe seed produces $s$-process 
distributions weighed towards Pb, with [Pb/Ba] ratios in some 
cases exceeding unity  and 0$<$[Ba/Sr]$<$1. 
%Increasing the value of M$_{\rm mix}$ in all the 2 \msun\ and 
%2.5 \msun\ models the [Pb/Ba] ratio decreases. This is 
%because the more heavy 
%elements are produced the lower becomes the neutron per Fe seed ratio as the 
%star evolves. 
In the massive 5.5 \msun\ model the only effect of including the 
$^{13}$C pocket is the increase in the production of Pb by a factor of 
$\simeq$50, so that this is the only massive AGB model where the final [Pb/Ba] 
ratio is positive. 
% AK - this sentence is not clear because regime 1 does *not* involve
% activation of the 13C neutron source!
%
%As for regime 1, any regime involving efficient activation 
%of the $^{13}$C neutron source produces [Ba/Eu] ratios close to unity.
% AK - do you mean:
Efficient activation of the $^{13}$C neutron source produces 
[Ba/Eu] ratios close to unity, which is typical of the $s$ process and
similar to what was found for Regime 1.
}

\item{{\it The $^{13}$C($\alpha$,n)$^{16}$O neutron source operates in convective 
conditions, with $^{13}$C produced by the inclusion of a $^{13}$C pocket (cases with  
M$_{\rm mix} \neq$ 0).} In this case the $^{13}$C in the $^{13}$C pocket does not 
completely burn radiatively and is ingested in the 
following TP. This regime characterises the first two or three $^{13}$C pockets in 
the lowest mass models, e.g., in the 1.25 \msun, and 1.5 \msun\ 
models the first three $^{13}$C pockets are engulfed in the 
following TPs while the abundance of $^{13}$C is still higher than 10$^{-4}$ (by number), and it 
is close to 10$^{-3}$ for the first $^{13}$C pocket. In the 1 \msun\ Stromlo model we 
find two TDU episodes and thus two $^{13}$C pockets are included (cases with M$_{\rm mix} 
\neq$ 0), both of which are ingested in the 
following TP while the abundance of $^{13}$C is still higher than 10$^{-4}$. The neutrons
are released over material in the whole intershell region and the $^{13}$C is mixed 
with the $^{14}$N poison from the $^{13}$C pocket and from the H-burning ashes. Thus, the 
number of neutrons per Fe seed is lower with respect to the case when \iso{13}C 
burns completely in radiative conditions. Consequently, the [Pb/Ba] ratio is also lower.} 
%In constrast to the 2 \msun\ models 
%described in the previous regime, in the Stromlo 1.5 \msun\ model, with 
%increasing the value of M$_{\rm mix}$ the [Pb/Ba] ratio increases. This is a direct 
%consequence of the $^{13}$C pocket burning convectively as in this case the 
%production of Pb builds up as a function of the number of neutrons convectively 
%released in the TPs. In the {\sc stars} 1.5 \msun\ model increasing the 
%value of M$_{\rm mix}$ decreases the [Pb/Ba] ratio instead, indicating than in this model the previous regime with $^{13}$C burning convectively is more important than in the corresponding 
%Stromlo model.}

\item{{\it The $^{13}$C($\alpha$,n)$^{16}$O neutron source operates in
convective conditions, with $^{13}$C produced by the ingestion of a small
number of protons in the TPs.} This event occurs in the first few TPs in
models with masses roughly less than 2.5 \msun. The convective TP reaches
at most the point in mass where the H mass fraction is 0.01 and we
never find a splitting of the convective TPs as is seen to occur at lower
metallicities and/or in other studies \citep[e.g.,][]{campbell08,cristallo09b,lau09,suda10}.
However, the modeling of the proton ingestion event strongly depends on
the treatment of convection and the convective mixing scheme employed. In
the Stromlo code we instantaneously mix convective regions at each
iteration, which is not appropriate in cases such as proton ingestion episodes when 
mixing and burning timescales are comparable.
In the {\sc stars} code we use a diffusive mixing scheme \citep{eggleton72},
which would be able to resolve the splitting of the convective region if
it occurred. On the other hand, the first 3D simulations of  AGB proton
ingestion episodes considered here suggest that an advective mixing scheme may be 
more appropriate than a diffusive mixing scheme
and that the splitting of the convective TP
does not occur \citep{stancliffe11}. With these caveats in mind, we find
that proton ingestion episodes can be expected to lead to favorable
conditions for $s$-process nucleosynthesis and to favour the
production of the $s$-process elements lighter than Pb, for the same reasons
related to Regime 3 discussed above.
While for models with mass higher than $\simeq$ 1.9 \msun\ the effect 
of these events is dwarfed by the presence of the 
$^{13}$C pocket, in the lower mass models the proton-ingestion events 
alone can result in significant production of $s$-process 
elements (cases with M$_{\rm mix}$=0).
The results are relatively low [Pb/Ba] ratios and the 
usual $s$-process [Ba/Eu] ratio near unity. The final [Ba/Sr] in these low-mass models depends 
on the detailed features of the proton-ingestion episodes, with ingestion of a smaller number of
protons favouring 
the production of Sr, as well as on the amount of $^{13}$C 
from the pocket that is left over to burn in the following TP (Regime 3). 
In some of the {\sc stars} models the final 
result is a negative [Ba/Sr] ratio.  
Production of the light element Na due to the $^{22}$Ne$+$p during the proton-ingestion events  
occurs in most of the low-mass models (cases with M$_{\rm mix}$=0).}

\end{enumerate}

As outlined above, the Stromlo and {\sc stars} models differ with regards to the masses 
at which the neutron-capture regimes are predominant. Thus, 
the relative distribution of the $s$-process elements (the [Ba/Sr] and [Pb/Ba] ratios)
may be different for models of the same stellar mass (compare, e.g., the distributions plotted
in Fig.~\ref{fig:heavyresults} for the Stromlo and {\sc stars} models of same mass 1.25 \msun.) 
The other main difference is that the 
absolute abundances with respect to Fe are typically smaller in the {\sc 
stars} than in the Stromlo models. 
This is because, as discussed in 
\S~\ref{sec:models}, the {\sc stars} models have a smaller number of TDU episodes due 
to the inclusion of realistic molecular opacities which
vary with the envelope composition and which result in a shorter
AGB lifetime. 
The same effect, though to a smaller extent, appears when considering the Stromlo 2 \msun\ 
and 2.5 \msun\ models computed with the inclusion of molecular opacities. 

We changed the initial abundances in most of our nucleosynthesis post-processing 
models according to the different choices indicated in \S~\ref{sec:nucmodels} 
and did not find any remarkable difference in the any of the final 
abundances reported in Tables~\ref{tab:resultsSTARS} and
\ref{tab:resultsMM} (as well as in Table\ref{tab:resultslight} discussed below). 
The only 
significant change occurred to [Eu/Fe] when we applied a significant $r$-process 
enhancement. This is obvious as Eu is a predominantly $r$-process element. 
For example, in the Stromlo 2 \msun\ model computed with M$_{\rm mix}$=0.002
\msun, using an $r$-process contribution to the initial composition of 
[$r$/Fe]=0.0, 0.4, 1.0, and 2.0 dex, we obtained final [Eu/Fe]=1.51, 1.53, 1.60, and 
2.08, respectively. In all cases, the [Ba/Fe] was unaffected, leading to final
[Ba/Eu]=0.91, 0.89, 0.82, and 0.39, for the four cases respectively.

\subsection{The light elements: C, N, O, F, Ne, Na, and Mg}
\label{sec:light}

\begin{figure} 
\begin{center}
%\epsscale{.80} 
%\plotone{BaEu_Full.ps} 
\includegraphics[scale=0.6,angle=0]{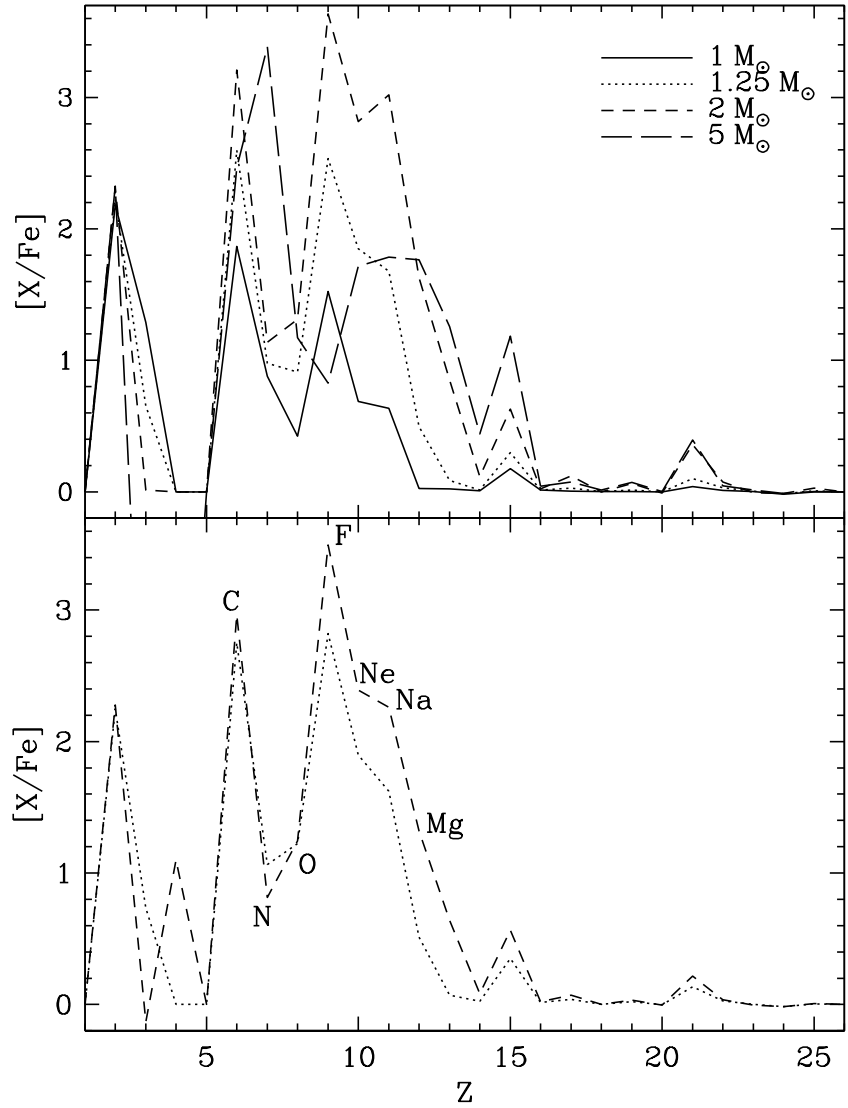} 
\caption{Same as Fig.~\ref{fig:heavyresults} but showing the [X/Fe]
ratios as a function of the atomic number Z for elements lighter than
Fe. Stromlo models are shown in the upper panel and {\sc stars} 
models in the lower panel.}
\label{fig:lightresults}
\end{center}
\end{figure}

The C, N, O, and F abundances in low-metallicity AGB stars have been already 
extensively presented by \citet{herwig04b}, \citet{karakas07b}, 
\citet{lugaro08a}, \citet{stancliffe09} and \citet{karakas10a} so here we 
only discuss them briefly. 
In Fig.~\ref{fig:lightresults} we present the final [X/Fe] for the elements 
lighter than Fe for the same selected models of Fig.~\ref{fig:heavyresults}. 
In Table~\ref{tab:resultslight} we present 
the final [X/Fe] for C, N, O, F, 
and Ne for all our models and the indicated choices of M$_{\rm mix}$.
These light elements mostly do not change when varying this free
parameter: the only exceptions are for F and Ne. For example, F
decreases by 0.41~dex and Ne by 0.45~dex in the Stromlo 1 \msun\ model when  
M$_{\rm mix}$=0
All our models are very rich in carbon and relatively poor in 
nitrogen, with typical [C/N] varying from $\simeq$ 1 dex, for the 0.9 and 1 \msun\ models, 
to $\simeq$ 2 dex for models up to 3 \msun\ (Fig.~\ref{fig:lightresults}). 
From this stellar mass and above, HBB 
is activated and the nitrogen abundance increases and quickly overcomes the abundance 
of carbon.  
Oxygen is produced in our models, ranging from 0.42 to 1.34 dex. A typical product 
of AGB nucleosynthesis is fluorine, which we find to be strongly produced in our models: 
[F/Fe] is comparable to [C/Fe] up to the stellar mass where HBB is
efficient. For higher masses fluorine is destroyed by proton captures. Neon is 
also a typical product of AGB stars, in the form of \iso{22}Ne, and its production 
factor varies in our models from $\sim$+1 to $\sim$+3 dex \citep{karakas03b}.
This element is observable in planetary nebulae and can be used
as a further probe of AGB nucleosynthesis. The Ne abundance was derived for the halo 
planetary nebula BoBn~1, which is also rich in C, N, and F, and can be
explained by low-metallicity AGB models of $\approx$ 1.5 \msun\ \citep{otsuka10}.
The production of F and Ne is primary in low-metallicity AGB 
stars because it is driven by the TDU of \iso{12}C. Our 2 \msun\ 
models produce very similar results to the model of \citet{cristallo09a}. 

The light elements Na and Mg require specific discussion because they depend on the 
abundance of \iso{22}Ne, which in turn depends on the amount of TDU and therefore on 
the stellar mass, on the formation of the \iso{13}C pocket, and on the overall 
neutron flux. Our predictions for these elements are shown in 
Fig.~\ref{fig:lightresults} for selected models and listed in 
Tables~\ref{tab:resultsSTARS} and \ref{tab:resultsMM} for all the models. Na is 
strongly produced in our models due to proton and neutron captures on \iso{22}Ne. For 
models without HBB, the increasing efficiency of the TDU with stellar mass means that 
the amount of primary \iso{22}Ne and Na in the envelope also increase with mass up to 
$\simeq$ 2 \msun; this is illustrated in Fig.~\ref{fig:lightresults}. In these 
low-mass models the \iso{23}Na is synthesized by proton captures on \iso{22}Ne during 
both the interpulse and convective TP \citep[see also][]{goriely00,cristallo09a}.
Neutron captures on \iso{22}Ne occur any time there are neutrons 
available given the large abundance of \iso{22}Ne present in the
intershell \citep[see also the discussions in][]{karakas10a,bisterzo10}. 
In models of masses higher than around 2 \msun, less Na is produced 
because the total amount of TDU as well as M$_{\rm mix}$ are smaller and 
because Na production by HBB is sensitive to mass. Na
production requires higher temperatures than CNO cycling and takes
place most efficiently at $\approx$~4 \msun; at higher masses the
temperature at the base of the envelope is so high that the Na is
effectively destroyed by proton captures.
At the other end of our mass range, in the lowest-mass models 
0.9 and 1 \msun, we also obtain a range of Na abundances, depending
on the features of the proton ingestion and the inclusion 
of the $^{13}$C pocket.
The 0.9 \msun\ and 1.0 \msun\ Stromlo models are the only models to
have [Na/Fe] significantly lower than [Ba/Fe]. In all the 
other models [Na/Fe] is comparable and even higher than [Ba/Fe]. 

Magnesium is also significantly produced in our models when the mass is greater than 
$\simeq$ 1.25 \msun. In the models with mass lower than $\simeq$ 2.5 \msun\ all the 
isotopes of Mg are produced via neutron captures in the \iso{13}C pocket 
starting on \iso{23}Na. 
For this reason the abundance of Mg follows the abundance of Na, keeping roughly to 
one order of magnitude lower than Na. In the models with mass higher than $\simeq$ 
2.5 \msun, the heavier isotopes of Mg, \iso{25}Mg and \iso{26}Mg, are also produced 
by $\alpha$-capture during the TP via the \iso{22}Ne+$\alpha$ reactions. HBB 
in the most massive models leads to activation of the Mg-Al chains, which produces 
\iso{26}Mg and \iso{26}Al and alters the composition of the He-shell prior to each
TP \citep{karakas03b}. In these higher mass models the abundance of Mg becomes 
comparable to, and even larger than, that of Na.

\begin{figure}
\begin{center}
%\epsscale{.80}
%\plotone{BaEu_Full.ps}
\includegraphics[scale=0.6,angle=0]{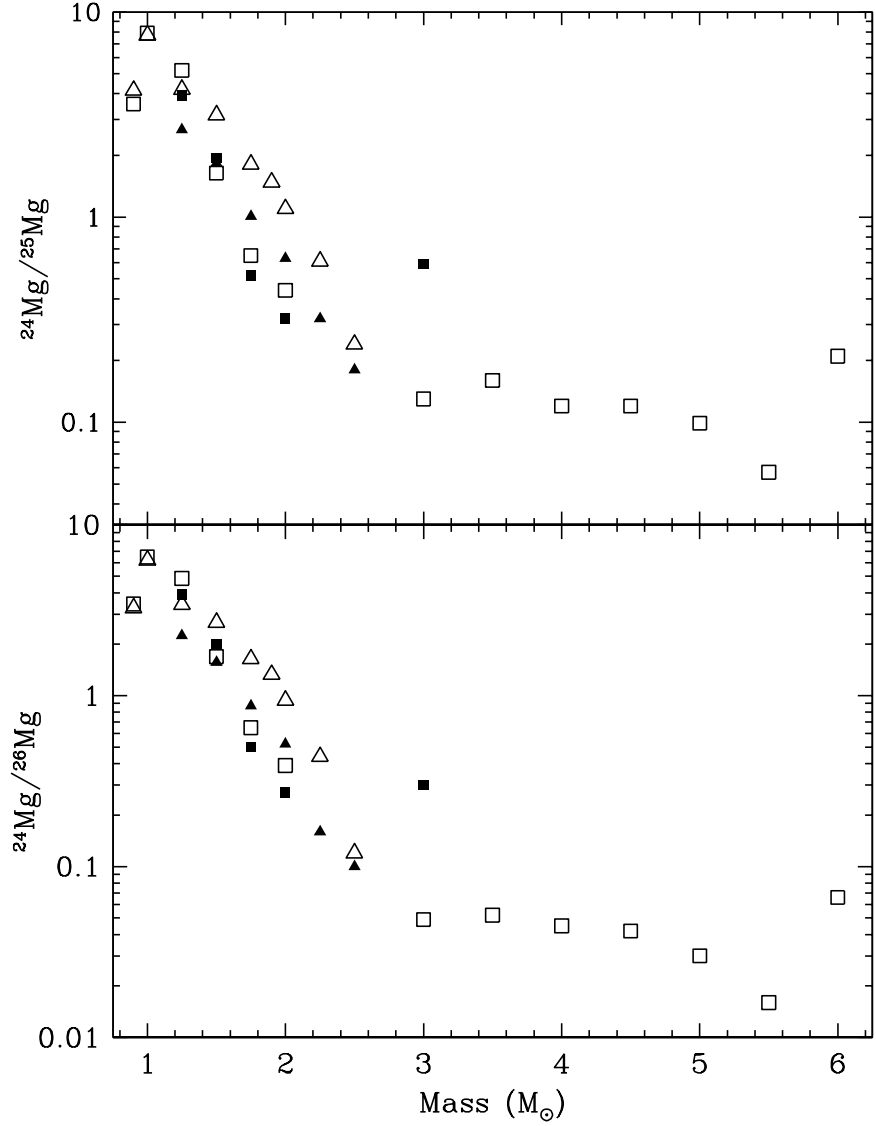}
\caption{Mg isotopic ratios (by number) as function of the initial stellar mass: 
open symbols represent Stromlo models, black symbols represent {\sc stars} models. Squares 
are for models with M$_{\rm mix}$=0 and triangles are for models with
M$_{\rm mix}$=0.002 \msun.}
\label{fig:mgiso}
\end{center}
\end{figure}

In Fig.~\ref{fig:mgiso} we show the Mg isotopic ratios at the
stellar surface at the end of the computed evolution. These 
ratios vary by up to three orders of magnitude with the stellar mass.
The derivation of the the Mg isotopic ratios
from the spectra of CEMP stars would represent very strong
constraints on the mass of the donor AGB star. We note 
that the models shown in Fig.~\ref{fig:mgiso} were computed using
the solar ratios of \iso{24}Mg/\iso{25}Mg=7.9 and
\iso{24}Mg/\iso{26}Mg=7.2.  In selected models with mass $\ge 1.5$ \msun\
we tested the effect of starting with ratios derived from the detailed 
Galactic chemical evolution models of \citet{kobayashi11}, i.e., 
\iso{24}Mg/\iso{25}Mg=158 and \iso{24}Mg/\iso{26}Mg=179. 
While this did not have any impact on the final ratios for the
masses considered, the final compositions of stars with masses lower
than 1.5 \msun\ may be more strongly effected by the initial
composition. This is because these models experience less dredge-up
and for this reason, the final Mg isotopic ratios are likely to be
higher than given here.

\subsection{Comparison with previous models}
\label{sec:comparemodels}

In Table~\ref{tab:comparemodels} we present a comparison of our results with those of 
\citet{cristallo09a} and \citet{bisterzo10} for the same element ratios as shown in 
Tables~\ref{tab:resultsSTARS} and \ref{tab:resultsMM}. The results from the {\sc stars} and 
Cristallo's 2 \msun\ models show very good agreement, both in the light (see 
Table~\ref{tab:resultslight}) and in the heavy elements. Both these models were computed with the 
inclusion of molecular opacities. The Stromlo 2 \msun\ models both with and without inclusion of 
realistic molecular opacities produces higher abundances of both light and heavy elements in AGB models,
with respect to Fe. 
Varying the mass of the \iso{13}C pocket can also lead to significant changes in the predicted
abundances of both light and heavy elements in AGB models. The model by 
\citet{cristallo09a} includes the \iso{13}C pocket using a self-consistent approach based on
convective overshooting motivated by hydrodynamical simulation of convective/radiative 
boundaries. These authors 
found that the extent in mass of the pocket decreases with the TP number. In 
our models instead we kept M$_{\rm mix}$ constant with the TP number. 
This explains why the 
abundances with respect to Fe by \citet{cristallo09a} are closer to those computed with the smaller 
M$_{\rm mix}$ value in our models. 
In any case, the ratios [Ba/Sr], [Ba/Eu], and [Pb/Ba] are very similar in all three sets of 
models. This comparison indicates that the $s$-process distributions obtained using our 
artificial treatment of the \iso{13}C pocket are in agreement with those obtained using the 
self-consistent treatment of \citet{cristallo09a}. 

\citet{bisterzo10} treated the neutron 
production in the \iso{13}C pocket as a free parameter and varied its efficiency with respect to 
their standard (ST) case, which reproduces the solar system $s$-process abundances at $Z=0.01$. 
This approach results in a larger allowed range of variations of the
resulting abundances. 
The results from the Stromlo, {\sc stars}, and Cristallo's
models are typically included within the range predicted by \citet{bisterzo10}. 

Overall, Table~\ref{tab:comparemodels} presents a relatively consistent set of 
results for neutron-capture nucleosynthesis in AGB stars of low metallicity. 
The main features of these 
models are the production of $s$-process elements up to $+$3 dex and a variety of 
$s$-process distributions, typically with 0$<$[Ba/Sr]$<$1.0, [Ba/Eu]$\simeq$0.9, and 
0$<$[Pb/Ba]$<$2.0. [Ba/Eu] reaches lower values only if there is not a 
large $s$-process enhancement, i.e., [Ba/Fe]$<$1.0. The ``radiative + convective'' 
model of \citet{goriely05} produces a heavy-element distribution in the range 
displayed in Table~\ref{tab:comparemodels}, with [Ba/Sr]$\simeq$0.9, 
[Ba/Eu]$\simeq$0.8, and [Pb/Ba]$\simeq$1.0, see their Fig.~6. The production of Na 
and Mg always increases with the stellar mass, something also seen in the 
\citet{bisterzo10} models. Their 1.3 \msun\ models produce lower Na than the 
1.25 \msun\ models presented here because these models do not include production 
of Na via proton captures.

Finally we find that, in spite of the fact that the mixing scheme strongly
affects the proton-ingestion episodes, our results for the $s$ process are
qualitatively the same as the results presented by \citet{cristallo09b} for
a 1.5 \msun\ star of metallicity half of that used in our models 
($Z = 5 \times 10^{-5}$) 
which also experience a proton-ingestion episode. From Fig.~6 of
\citet{cristallo09b} we see that after the first TDU episode, which records
the effect of the proton-ingestion event, 
the $s$ process favours the production of the lighter elements, just like our models. 
The final
[Ba/Sr]$\simeq$0.40, [Ba/Eu]$\simeq$0.95, and [Pb/Ba]$\simeq$0.85 are
close to those obtained in our 1.25 \msun\ and 1.5 \msun\ models.

\section{Comparison with the abundances of CEMP-$s$ and CEMP-$s/r$ 
stars}\label{sec:compare}

\subsection{[Ba/Fe] versus [Eu/Fe]}
\label{sec:baeu}

\begin{figure}
%\epsscale{.80}
%\plotone{BaEu_Full.ps}
\includegraphics[scale=0.6,angle=270]{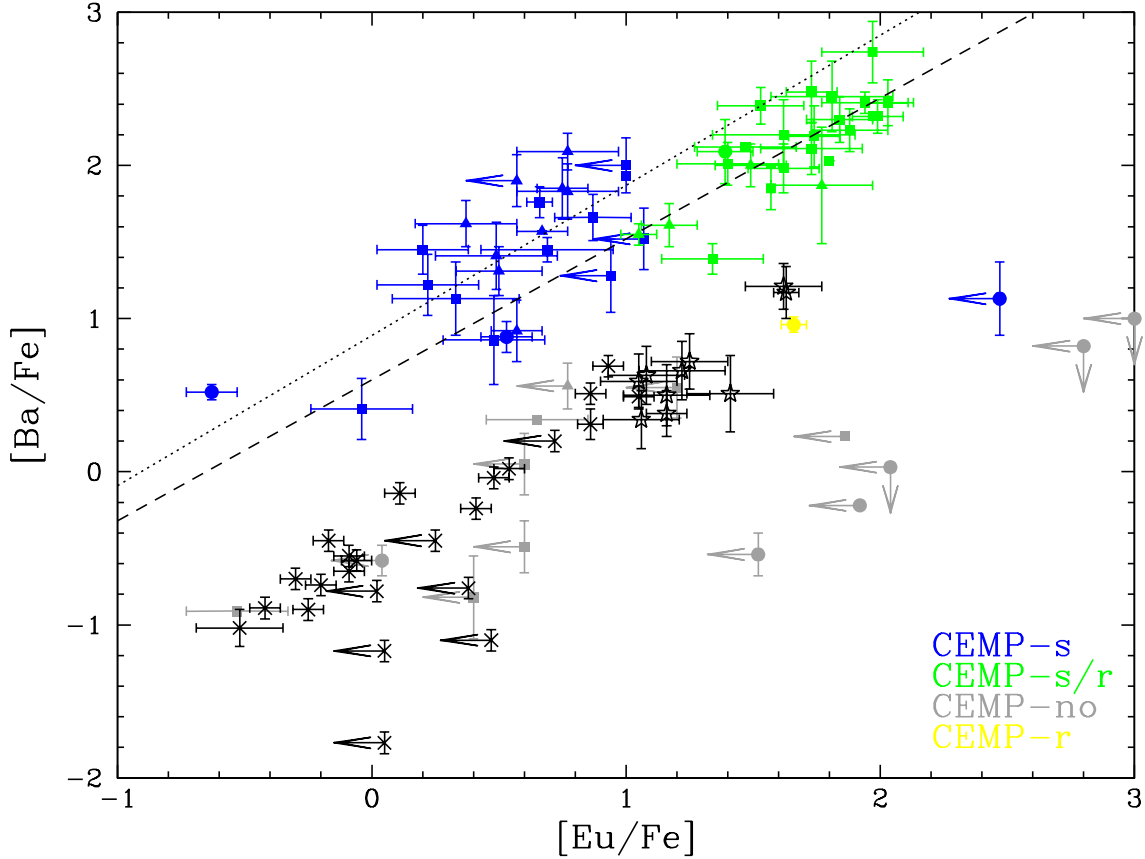}
\caption{Overview of the observational data for CEMP  
stars (colour-coded as in the legend) and $r$I (black crosses) and $r$II (open star
symbols) stars following the data and classification of
\citet{masseron10} except that, for sake of simplicity, we have included the   
CEMP-low-$s$ in the CEMP-$s$ group. 
For CEMP stars, triangles represent stars with [Fe/H] $> -2$, 
squares represent stars with $-3<$[Fe/H]$<-2$, and circles represent 
stars with [Fe/H]$< -3$.
The correlation lines going through 
CEMP-$s$ (short-dashed line corresponding to [Ba/Fe] = 0.98 [Eu/Fe] + 0.89, 
with a correlation coefficient r=0.58)
and CEMP-$s/r$ (long-dashed line corresponding to [Ba/Fe] = 0.92 [Eu/Fe]+ 0.60, 
with correlation coefficient r=0.59) are also plotted. 
(A color version of this figure is available in the online
journal.)}
\label{fig:baeudata}
\end{figure}

Figure~\ref{fig:baeudata} presents an overview of the observational data for CEMP stars 
and carbon-normal metal-poor stars of the $r$I and $r$II subclasses following the data and 
classification of \citet{masseron10}. In this and all the following figures we used 
different symbols for stars in different metallicity ranges, as indicated in the caption. 
There are no noticeable differences in the distribution of any of the plotted abundances 
when changing the metallicity, the only noticeable properties being that there are more 
CEMP-$s$ than CEMP-$s/r$ stars with [Fe/H]$>-2$, and more CEMP-no
stars with [Fe/H]$<-3$ than in the other CEMP star classes.
The different classes of CEMP stars 
appear to be fairly well distinguished in this plot. 
Most of the CEMP-no Eu data are upper 
limits, and the only CEMP-$r$II star, CS\,22892-052, appears 
indistinguishable from the rest of the $r$II group in terms of its
heavy element composition.
 
The two groups of CEMP-$s$ and CEMP-$s/r$ appear to be distinct, with 
CEMP-$s/r$ having higher [Eu/Fe] ratio in absolute terms, but also higher [Ba/Fe] on average. 
In Fig.~\ref{fig:baeudata} we plot the correlation lines of the two groups. The similar 
slope, around 1, indicates that for both classes the linear dependence would be preserved in a 
non-logarithmic scale. What counts here is the y-intercept, which corresponds to [Ba/Eu] and thus 
defines the slope of the linear correlation between the two variables in a non-logarithmic scale:
in the case of CEMP-$s$ stars
the process responsible for their compositions produces on average 
of around 400 times more Ba than Eu, with a spread of a factor of about 2;
in the case of CEMP-$s/r$ stars, instead, 
the process responsible for their composition produces on average 
of around 200 times more Ba than Eu, with a spread of a factor of about 2. Not surprisingly, the 
$r$ stars also sit on a correlation line with a slope of around 1 and y-intercept of around $-0.6$, 
corresponding to the solar-system $r$-process Ba/Eu ratio of about 10. 
In the case of CEMP-$s$ the [Ba/Eu] value of +0.89 agrees very well with the typical [Ba/Eu] ratios 
reported in Tables~\ref{tab:resultsSTARS}, \ref{tab:resultsMM}, and \ref{tab:comparemodels}, 
while in the case of CEMP-$s/r$, the value of +0.6 does not agree with any AGB models 
producing [Ba/Fe]$>$1.

The three CEMP-$s/r$ with the lowest Ba abundance in their 
group may belong instead to the CEMP-$s$ group. 
This seems likely for BS~16080-175, with [Ba/Fe]=1.55 and [Eu/Fe]=1.05, which can be 
reclassified as CEMP-$s$ within the 0.07 dex uncertainties. The other two outliers 
BS~17436-058, with [Ba/Fe]=1.60 and [Eu/Fe]=1.17, and HD~187861, with [Ba/Fe]=1.39 and 
[Eu/Fe]=1.34, may be reclassified as CEMP-$s$ within 2$\sigma$ of their 0.11 and 0.20 dex 
[Eu/Fe] uncertainties. However, we note that [La/Fe]=1.73$\pm$0.2 in HD~187861 
indicates that it should be re-classified
as a CEMP-$s/r$ star. More observations and analysis of these 
stars will help to clarify
their status. 

Another interesting case is that of CS\,30322-023 ([Ba/Fe]=0.52 and 
[Eu/Fe]=$-$0.63), which has been previously classified as a CEMP-$s$ \citep{aoki07}, and was 
included in the CEMP-low-$s$ class by \citet{masseron10}. 
This star has the highest [N/C] $\simeq +2$ and the lowest, and only 
negative, [Eu/Fe] ratio in the sample, as well as [Fe/H]$\simeq -$3.3, at the 
lower end of the CEMP-$s$ range. Another three stars have [N/C] $> 1$: 
CS\,22949-037, CS\,22960-053, and CS\,29528-041, which all have [Fe/H] similar to 
CS\,30322-023 and are classified as CEMP-no by \citet{masseron10}. However, CS\,22960-053 and 
CS\,29528-041 have [Ba/Fe]$\simeq$0.9 
but no Eu observations, so they may be 
belong to the CEMP-low-$s$ class \citep[as defined by][]{masseron10}, together with CS\,30322-023.
In fact, \citet{aoki07} classified both CS\,22960-053 and CS\,30322-023
as CEMP-$s$ stars following \citet{ryan05}. 

\begin{figure}
\begin{center}
%\epsscale{.80}
%\plotone{BaEu_WithModels.ps}
\includegraphics[scale=0.7]{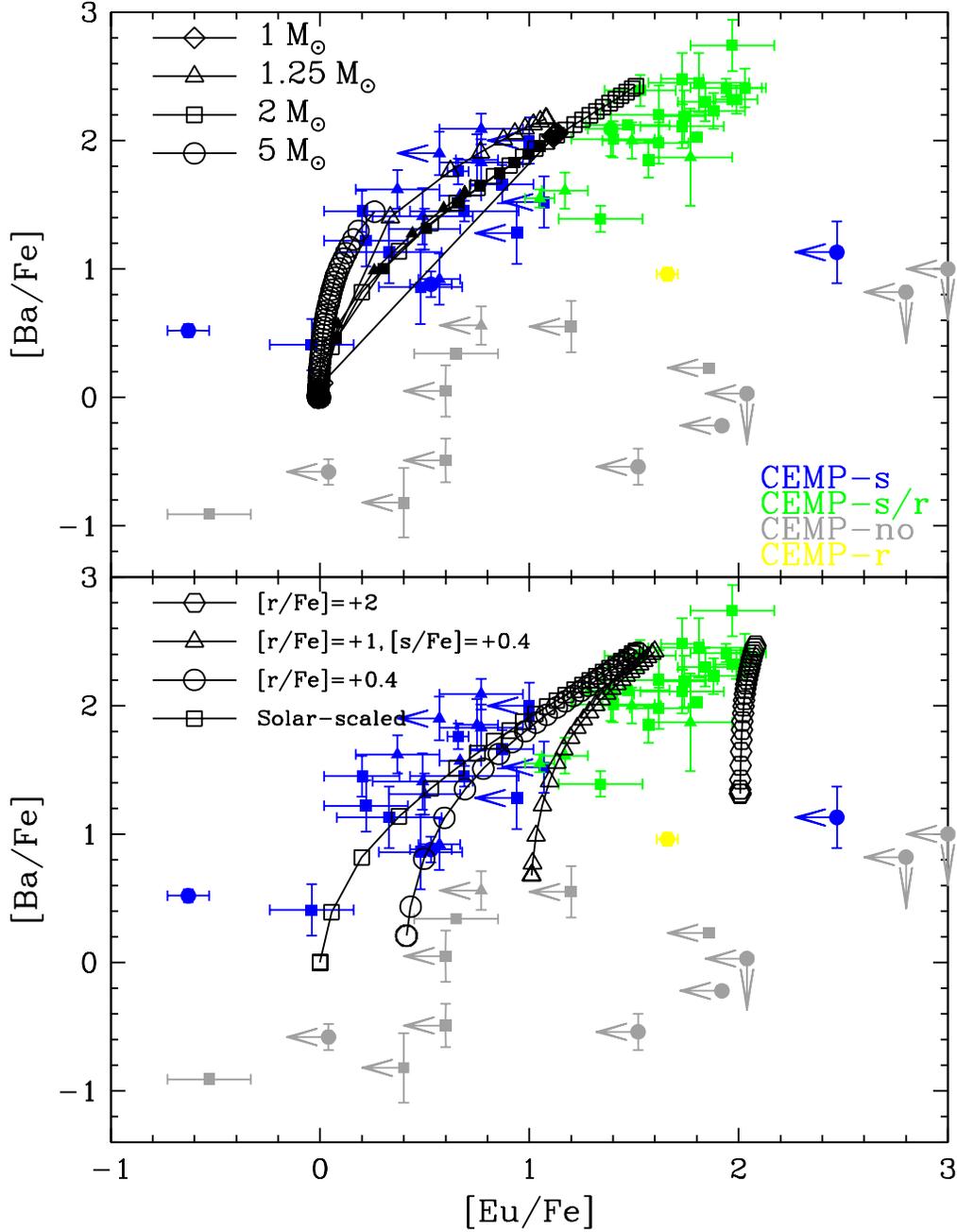}
\caption{A selection of our model predictions is plotted together with the 
CEMP data of Fig.~\ref{fig:baeudata}. The 
predictions lines from our models represent the evolution of the stellar surface 
composition of the primary AGB stars and each symbol on the lines represents a TP. 
Top panel: open symbols refer to the Stromlo models and black 
symbols refer to the {\sc stars} models. Stellar masses are indicated in the legend.
Bottom panel: open symbols refer to Stromlo models of 2 \msun\ and different 
initial compositions as indicated in the legend. (A 
color version of this figure is available in the online journal.)}
\label{fig:baeumodels}
\end{center}
\end{figure}

Figure~\ref{fig:baeumodels} includes a selection of our model predictions together 
with the data of Fig.~\ref{fig:baeudata}. In this as in all the following figures
the solid lines are a visual aid connecting the predicted envelope composition 
after each TDU episode. These 
prediction lines are not straight because the composition of the $s$-process 
material in the He intershell evolves with time and because the variables are 
shown on a logarithmic scale. The process of transferring and mixing AGB material onto 
the companion star would result in mixing lines connecting the AGB 
composition with the initial composition. These mixing lines closely follow the 
plotted AGB evolution lines and are not shown in the figure.
We include in the figure the same stellar models plotted in Figs.~\ref{fig:heavyresults} and
\ref{fig:lightresults}, representing cases where the neutron-capture nucleosynthesis occurs in the 
different regimes described in \S~\ref{sec:results}. 
All the plotted models start from a 
solar composition, [Ba/Fe]=[Eu/Fe]=0, except for the three 2 \msun\ models from the 
Stromlo set with different initial composition: [$r$/Fe]=$+$0.4, [$r$/Fe]=$+$1.0, and [$r$/Fe]=$+$2.0 
included in the lower panel of Fig.~\ref{fig:baeumodels}. 

All the models produce both Ba and Eu. The predictions lines follow the trend of the 
CEMP-$s$ group. A small spread in the initial {\it r}-process composition up to 
[$r$/Fe]=$\pm$0.4 dex, which is within the composition observed in 
field stars of the corresponding metallicity \citep[see, e.g., Fig.~2 of][]{bisterzo11a},
is enough to cover the data.
CS\,30322-023 indicates an initial negative [$r$/Fe]$\approx -$0.6, 
which would also be within the observed range at the corresponding metallicity [Fe/H]=$-$3.3.
On the other hand, AGB models do not produce the high [Eu/Fe] 
observed in CEMP-$s/r$ stars. 

One solution to 
this problem may be to increase the initial [$r$/Fe] abundance by up to $+$2
dex under the assumption that the $r$-process enrichment is primordial in these 
systems (bottom panel of Fig.~\ref{fig:baeumodels}). 
This primordial enrichment can arise by assuming that the 
molecular clouds where CEMP-$s/r$ 
stars formed were polluted by $r$-process elements from a
nearby SNII.  This scenario 
provides a solution for the highest measured [Eu/Fe] abundances in CEMP-$s/r$ 
stars as discussed in detail by \citet{bisterzo11a}. However, it comes with three 
problems, some of which have 
been discussed previously \citep[see e.g.][]{jonsell06,lugaro09}. 
(1) The initial [$r$/Fe] value does not affect the final [Ba/Fe] value, which remains 
constant, hence the linear correlation observed between the Eu and Ba enhancements in the 
CEMP-$s/r$ sample is not reproduced. 
(2) The number of $r$II stars (around 10) is smaller than the 
number of CEMP-$s/r$ stars (about 30), so, it seems unlikely that CEMP-$s/r$ stars evolved 
from $r$II stars in binary systems, because these should make up a fraction of all $r$II 
stars. This does not appear to be an observational bias since stellar surveys such as 
the Hamburg/ESO R-process-Enhanced Stars (HERES) have specifically targeted $r$I and $r$II 
stars. It is also difficult to invoke supernova triggering as the cause for 
the formation of binaries: though early work indicated this link possible
\citep{vanhala98}, it is now understood that formation of binaries
occurs very naturally whenever a turbulent velocity field is present in 
a molecular cloud \citep{bate09}. Furthermore the whole concept of
supernova-triggered star formation is still controversial, both
observationally and theoretically \citep{leao09,elmegreen11}.
(3) The metallicity distribution is different for the two samples: 
[Fe/H] is centered at $\simeq -$2.5 for CEMP-$s/r$ stars and at 
$\simeq -$2.8 for $r$II stars.

\subsection{[ls/hs] versus [Mg/hs]}

In this and the following subsections we compare our model predictions to the observations using 
``intrinsic'' indicators, i.e., elemental ratios that include only elements that are produced in 
AGB stars. Different from the [Ba/Fe] and [Eu/Fe] ratios considered above, where Fe is not 
produced in AGB models, the ``intrinsic'' ratios move away from their initial solar values towards 
their $s$-process values already after a small number of TPs (though this is less true for the 5 
\msun\ model where the dilution in the envelope is much larger than in the other models). For the 
CEMP stars, this means that the intrinsic ratios are close to the AGB values even for relatively 
large dilutions of the AGB material onto the CEMP star. Hence,
intrinsic ratios are, in a first 
approximation, independent of the TDU, the mass loss, the stellar lifetime, and the accretion and 
mixing processes on the binary companion. Instead, they mostly constrain the nucleosynthesis 
occurring in the deep layers of the star and in our context, the neutron source and the 
neutron flux for the $s$ process. 

We also note that we consider here ratios of elements that are all
significantly produced in AGB stars, so that variations in the initial
composition of these elements due to the chemical evolution of the
Galaxy do not have a significant effect on the AGB predictions. Using
the initial composition from \citet{kobayashi11} (i.e., [F/Fe], [Na/Fe],
and [Mg/Fe] equal to $-$1.5, $-$0.2, and $+$0.3 dex, respectively) did    
not lead to modifications of the results plotted and discussed here in
selected models with mass $\ge 1.5$ \msun. Only the [Mg/Fe] ratio of
stars with masses lower than 1.5 \msun\ may be more strongly effected by
the initial composition because these models experience less dredge-up.
For these models, this effect would at most shift the [Mg/hs] ratios by 
$+$0.3 dex.

\begin{figure}
%\epsscale{.80}
%\plotone{LshsvMghs_WithModels.ps}
\includegraphics[scale=0.6,angle=270]{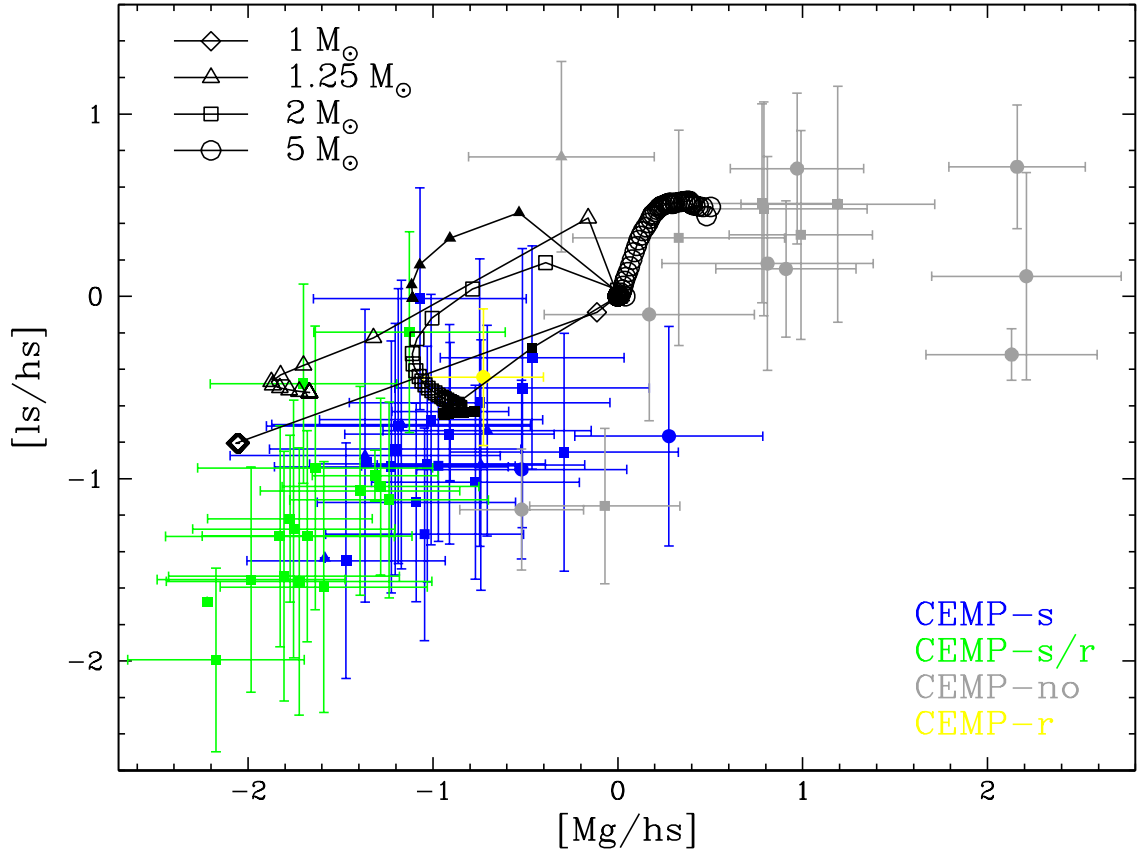}
\caption{Same as Fig.~\ref{fig:baeumodels} but for [ls/hs] versus [Mg/hs].
For data on the ls elements we searched the SAGA database \citep{suda08}.
(A color version of this figure is available in the online journal.)}
\label{fig:lshsmghs}
\end{figure}

Figure~\ref{fig:lshsmghs} shows the [ls/hs] and [Mg/hs] observed in CEMP-$s$ and CEMP-$s/r$ 
stars together with our model predictions. The dispersion of the data is quite large, 
however, it clearly shows the presence of a linear correlation with slope $\simeq$ 1. 
The y-intercept $\simeq$ 0 
indicates that this correlation is most likely due to to variations in the amount of the 
hs elements, Ba, La, and Ce. As noted in previous studies 
\citep[see, e.g., Fig.~9 and Fig.~6 of][respectively]{jonsell06,bisterzo11a}, the CEMP-$s/r$ 
stars are separated from the CEMP-$s$ in 
that they show [ls/hs] lower by at least 0.5 dex. Here we show that
CEMP-$s/r$ stars also show lower [Mg/hs] than CEMP-$s$.
It seems obvious to ascribe these differences to the fact that 
CEMP-$s/r$ stars have, on average, higher hs abundances than CEMP-$s$ 
(Fig.~\ref{fig:baeumodels}). 

All our models produce [ls/hs]$>-$1. This derives from the basic 
way in which the $s$ process operates \citep[see also][]{busso01}.
As the neutron flux increases, first the ls 
elements, then the hs elements, and finally Pb are produced. As the flux goes through 
these peaks the abundances reach their equilibrium values, above which it is not possible to 
increase them. 
This is similar to the situation of buckets connected to each other. When 
water is poured into the first bucket, the bucket fills up. Once it is full, the water 
starts moving into the second bucket and so on. The relative amount of water in the 
two buckets is only determined by their size (the neutron-capture cross sections in the 
case of the ls and hs elements) and not by how much more water is poured into the 
system\footnote{To finish with this simile, we note that the third bucket, corresponding 
to the Pb peak in the $s$ process, would have an infinite size.}.
Hence, the minimum [ls/hs] ratio is the value expected when the $s$ process is in equilibrium 
through the first and the second $s$-process peaks (with 
$\sigma_{\rm A} N_{\rm A} = constant$).
This [ls/hs] ratio is roughly equal to $-1$ because   
the neutron-capture cross section and the solar abundance of, e.g., 
$^{138}$Ba are about 1.5 and 6 times lower (respectively) than those of, e.g., $^{88}$Sr.
On the other hand, the [Pb/hs] ratio cannot reach an equilibrium value because Pb is at 
the end of the $s$-process chain. The Pb abundance 
can continue to grow indefinitely as more
neutrons are made available.

Unless we ascribe the overall mismatch to the large error bars, which still would not explain why 
CEMP-$s/r$ stars are separated from CEMP-$s$ stars in terms of both [ls/hs] and [Mg/hs],
it appears that a fraction of CEMP-$s/r$ stars do not 
carry a typical pure $s$-process signature in the relative abundances of their $s$-process 
elements. Together with Fig.~\ref{fig:baeumodels}, 
Fig.~\ref{fig:lshsmghs} demonstrates that (i) CEMP-$s/r$ stars have the highest 
abundances of hs elements, and thus the lowest [ls/hs] and [Mg/hs] down to 
$-$2\footnote{Specifically, 
the CEMP-$s/r$ star with [ls/hs]$\simeq -2$ is HE~0212-0557 observed by \citet{cohen06}. 
The hs value of this stars is based on values of [Ba/Fe], [Ce/Fe], and 
[La/Fe] all very similar to each other, while ls is based on 
[Y/Fe]=0.55, derived from three lines, and [Sr/Fe]=$-$0.05, derived from one line. Removing Sr
form the calculation of ls still results in a very low [ls/hs]=$-$1.69.}, and (ii) our 
$s$-process models can match the observations well for CEMP-$s$ stars, but cannot match those 
of CEMP-$s/r$ stars, not only in terms of their Eu abundance, but also in terms of 
their ls and hs abundances. 

Focusing on the CEMP-$s$ stars only, they appear to be overall better
matched by the models with the \iso{13}C neutron source burning radiatively.
The 5 \msun\ 
model, which could not be ruled out as a match to the observations on the basis of 
the Ba and Eu enhancements (Fig.~\ref{fig:baeumodels}), produces too much Mg and ls 
with respect to hs to be able to match any of the CEMP-$s$ and CEMP-$s/r$ data. 
Inspection of the results for the 5.5 \msun\ model in Table~\ref{tab:resultsMM} shows 
that inclusion of the \iso{13}C pocket does not solve this problem. 
On the other hand, the composition of CS\,30322-023, the only CEMP-$s$ with 
[Mg/hs]$>$0, is compatible with the models of 3 \msun\ and 5.5 \msun\ computed with the 
inclusion of the \iso{13}C pocket.

\subsection{[Pb/hs] versus [Mg/hs]}

\begin{figure}
%\epsscale{.80}
%\plotone{PbhsvNahs_WithModels.ps}
\includegraphics[scale=0.6,angle=270]{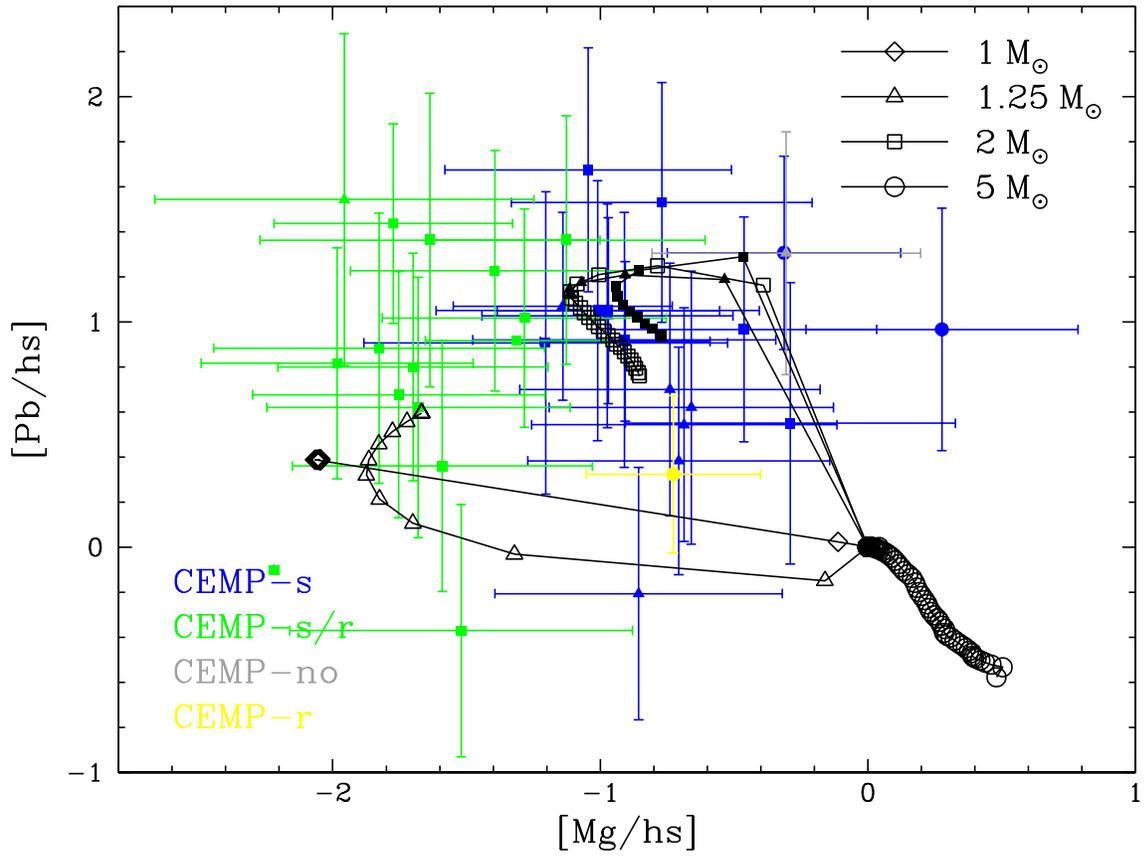}
\caption{Same as Fig.~\ref{fig:baeumodels} but for [Pb/hs] versus [Mg/hs].
(A color version of this figure is available in the online
journal.)}
\label{fig:pbhsmghs}
\end{figure}

Figure~\ref{fig:pbhsmghs} presents the [Pb/hs] and [Mg/hs] observed in CEMP-$s$ and 
CEMP-$s/r$ stars together with our model predictions. The dispersion of the data is large 
and there is no obvious correlation between the two ratios. While the two classes are 
different in all the other abundance ratios considered so far, 
the distribution of [Pb/hs] in CEMP-$s$ and CEMP-$s/r$ stars covers the same range.
The models with the \iso{13}C 
neutron source burning radiatively provide a good match to CEMP-$s$ stars, where the spread 
in the [Pb/hs] ratio can be explained by varying the TDU number or, equivalently, the 
initial stellar mass \citep[see also][]{bonacic07}.  
The models where the \iso{13}C 
neutron source burns convectively can provide a match for the lower end of the [Pb/hs] ratios 
observed in CEMP-$s/r$ stars, 
however, as shown in the previous section, they do not match the [ls/hs] values of 
CEMP-$s/r$. Even though the error bars are large, CEMP-$s/r$ stars with high [Pb/hs] 
ratios are not matched by any models. 

We note that the 5 \msun\ model does not cover any 
of the data points. As in the case of the [ls/hs] versus [Mg/hs] the composition 
of CS\,30322-023 can be matched by 
intermediate-mass AGB models (e.g., 3 \msun\ and 5.5 \msun) if a \iso{13}C 
pocket is included.

\subsection{[ls/hs] and [Pb/hs] versus [Na/hs] and [F/hs]}

\begin{figure}
\begin{center}
\includegraphics[scale=0.7]{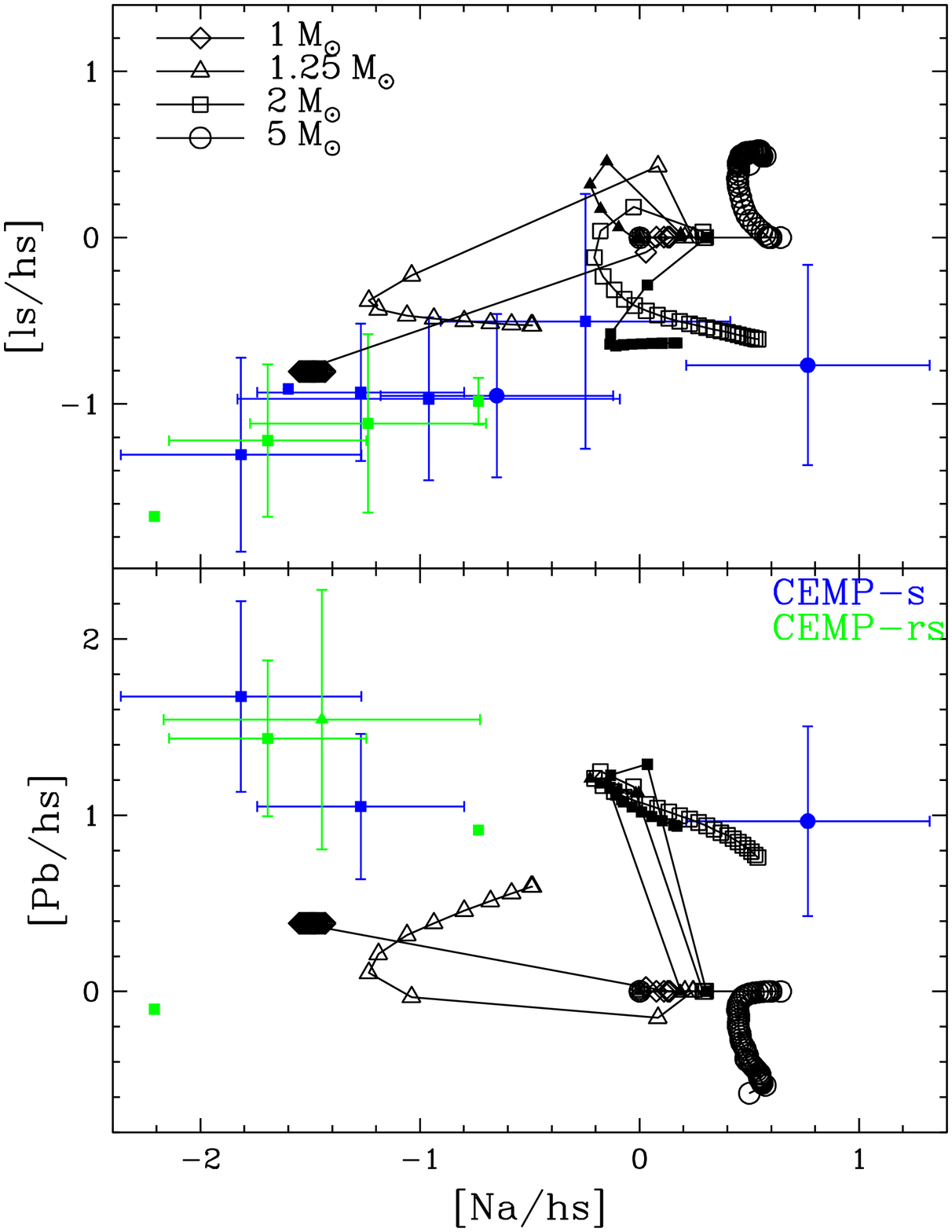}
\caption{Same as Fig.~\ref{fig:baeumodels} but for 
[ls/hs] versus [Na/hs] (top panel) and [Pb/hs] versus [Na/hs] (bottom panel).
Data for Na are from \citet{lucatello11}. 
(A color version of this figure is available in the online
journal.)}
\label{fig:na}
\end{center}
\end{figure}

\begin{figure}
\begin{center}
\includegraphics[scale=0.7]{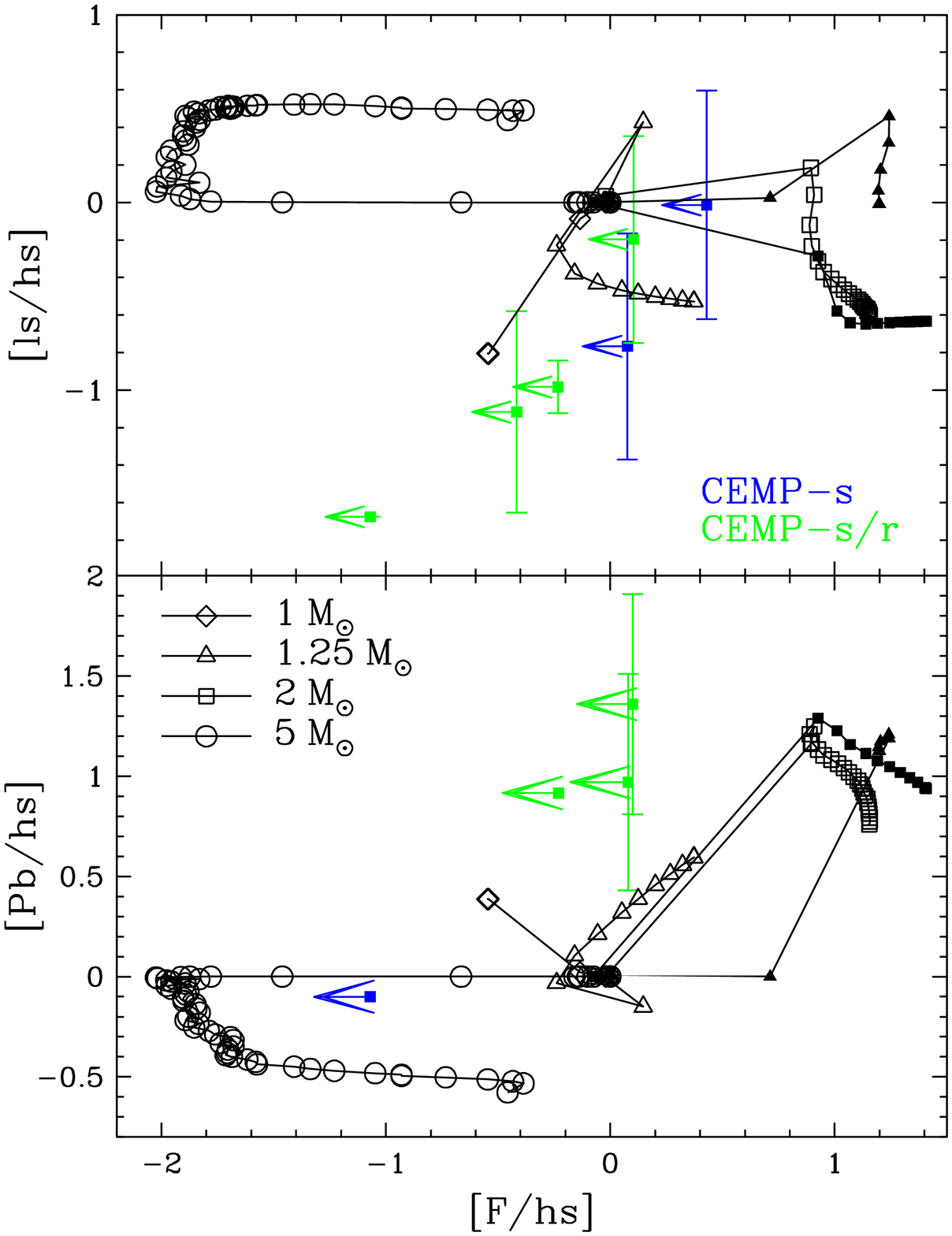}
\caption{Same as Fig.~\ref{fig:baeumodels} but for 
[ls/hs] versus [F/hs] (top panel) and [Pb/hs] versus [F/hs] (bottom panel).
Data for F are from \citet{lucatello11}. 
(A color version of this figure is available in the online
journal.)}
\label{fig:f}
\end{center}
\end{figure}

Figures~\ref{fig:na} and \ref{fig:f} present 
the [Na/hs] and [F/hs] ratios plotted versus the intrinsic $s$-process 
indicators, [ls/hs] and [Pb/hs].
There are not enough data to be able to determine if CEMP-$s$ and CEMP-$s/r$
stars are distinguished in their [Na/hs] and [F/hs] abundance ratios.
The models where the \iso{13}C neutron source burns radiatively, which 
provide a good match to the overall composition of CEMP-$s$ stars in terms of their 
[Mg/hs], [ls/hs], and [Pb/hs] compositions, typically produce too much 
Na and F with respect to the 
hs elements to match any of the observational data. One exception is the case of 
CS\,30322-023, the only CEMP-$s$ with [Na/hs]$>$0:  
this star is again distinguished from the rest of the CEMP-$s$ 
stars and as mentioned above may carry the signature of a more massive AGB star. 
A better overall match to [Na/hs] and [F/hs] is provided by the
low-mass Stromlo models where the \iso{13}C neutron source burns
convectively, however, these models do not cover the corresponding [Pb/hs] (and [ls/hs]) 
ratios. 

As discussed in \S~\ref{sec:light} the predicted abundance of Na is made by 
both neutron and proton captures on \iso{22}Ne. The fraction made by neutron 
captures is linked to the production of Mg, and the models where the \iso{13}C neutron 
source burns radiatively provide a good match to the 
[Mg/hs] observed in CEMP-$s$. The fraction made via proton captures, on the other hand, is 
disconnected from the production of Mg, but depends on the details of the proton-ingestion 
episodes, the profile of 
protons introduced to make the \iso{13}C pocket, and the proton-capture reaction rates
of $^{22}$Ne and $^{23}$Na \citep[which are uncertain, see][]{iliadis10}.
All these uncertainties need to be tested to see if 
it is possible to match the observed 
[Na/hs] and [Mg/hs] ratios, which indicate that the production of Na via proton captures 
should be minimised. During the proton diffusion leading to the formation of the 
\iso{13}C pocket, a proton profile weighed towards a low proton abundance
\citep[e.g., the ``fast'' profile in Fig.~10 of][]{goriely00} may contribute to the solution 
of this problem. This possibility 
will be investigated in detail in a forthcoming study and has the potential to 
provide a strong constraint on the formation of the 
\iso{13}C pocket.
A lower value of the $^{22}$Ne(p,$\gamma$)$^{23}$Na reaction rate or a higher
value of the $^{23}$Na(p,$\alpha$)$^{20}$Ne reaction rate could 
also help lower the predicted Na abundances.

The predicted [F/hs] ratios are also higher than the observed upper 
limits. The operation of some form of deep mixing of the envelope material of the 
AGB companion star down to regions where H burning occurs, which is also invoked 
to explain the observed high N enhancements and low \iso{12}C/\iso{13}C ratios 
\citep[see, e.g.,][]{stancliffe10} may contribute to decreasing the F abundance 
in these stars. Further investigations are required to test this idea. Note that in the 
5 \msun\ model with HBB, F is strongly depleted by proton
captures.

\section{Summary and conclusions}\label{sec:conclusions}

We have presented a large set of stellar models at metallicity [Fe/H]=$-$2.3 from 
two different evolutionary codes for which we have computed detailed $s$-process 
nucleosynthesis. For the first time we have presented abundance predictions for 
the elements heavier than iron from stars with masses lower than 1.3 \msun\ and 
above 2 \msun\ at this metallicity. The most up-to-date models including the effect of molecular 
opacities that vary with the envelope composition
during the AGB evolution experience less TPs, less TDU episodes, and produce 
overall lower enhancements of the $s$-process elements with respect to Fe.

We have found that there are four regimes of neutron-capture processes and the 
dominant regime depends on the initial stellar mass: for the highest mass models 
the \iso{22}Ne neutron source is dominant and the inclusion of a \iso{13}C pocket where \iso{13}C 
burns radiatively has an effect only on the production of Pb; 
for models of mass $<$ 3.5 \msun, 
\iso{13}C is the main neutron source. For masses $\simeq$ 2 \msun, \iso{13}C burns 
radiatively in the \iso{13}C pocket during the interpulse periods. For lower 
masses, \iso{13}C also burns convectively as the first few \iso{13}C pocket are 
engulfed in the following TPs before \iso{13}C is completely burnt in the interpulse
period. For the lowest 
masses, proton-ingestion episodes associated with the first few TPs also produce 
\iso{13}C, which burns convectively inside the TPs. Neutrons released
during convective TPs produce lower [Pb/hs] ratios than neutrons
released in the \iso{13}C pocket under radiative conditions, and
higher [ls/hs] ratios if a smaller number of protons are ingested.
It should be kept in mind that proton ingestion episodes are extremely
sensitive to the convective mixing scheme adopted. Our
models provide some insight into the effect of these episodes on the
$s$-process distribution, and are in agreement with the results presented by
\citet{cristallo09b}. However, more detailed 3D models of the
kind presented by \citet{stancliffe11} are needed to accurately
ascertain the effect of proton-ingestion episodes on the $s$-process in AGB stars.
The production of Mg and Na increases with the stellar mass as there is 
more primary \iso{22}Ne seed available.

From comparison with the observational data-set of CEMP stars compiled by \citet{masseron10} 
we have derived that our AGB models, especially those with radiative \iso{13}C burning, 
present a very good match to the composition of CEMP-$s$ stars. On the other hand, they 
cannot match the compositions of CEMP-$s/r$ stars, since they cannot produce the 
observed Eu enhancements up to one order of magnitude higher than in CEMP-$s$ stars. 
If the $s$ process and $r$ process components, making Ba and Eu, respectively, in CEMP-$s/r$ stars 
are completely independent, as in the scenario where the 
$r$ enrichment is primordial, it is difficult to explain why the observed Ba and Eu are 
correlated. This is because for any given initial [$r$/Fe]
abundance pattern imposed, the AGB models produce the same final
[Ba/Fe] abundance. Since this conclusion relies on the observed
correlation between Ba and Eu, it is  urgent to confirm and establish 
this correlation using a larger number of CEMP-$s/r$ star
abundances derived from high-resolution spectroscopy and
for a larger number of hs elements.

Not only do CEMP-$s/r$ and CEMP-$s$ stars differ in that CEMP-$s/r$ stars have more Eu than CEMP-$s$ stars in absolute terms, but they also differ in that CEMP-$s/r$ stars have on average more Ba 
(and other hs elements) than CEMP-$s$ stars. This behaviour is not readily explained 
if the $s$-process component in CEMP-$s$ and CEMP-$s/r$ stars come from the same 
AGB star sources. Furthermore, the higher Ba (hs) abundances are not 
accompanied by higher abundances in Mg and in the ls elements, as it is expected in 
our models. In fact, our AGB models always produce [ls/hs]$>-1$, which is too high to match 
the distribution of the $s$-process elements in CEMP-$s/r$ with [ls/hs] down to $\simeq -2$. 
On the other hand the higher Ba abundances 
are accompanied by higher Pb abundances, as expected in the $s$ process.
In summary, the composition 
of CEMP-$s/r$ stars is characterised by large
enhancements of the hs elements and of Eu (correlated with Ba: $\simeq 1/200$ Ba), 
as well as of Pb, but not of the ls elements. 
It remains to be seen if these compositions could be achieved 
by a ``$s/r$'' neutron-capture process, which may be a single neutron-capture 
process with features in-between or a superposition of the $s$ and $r$ processes.
Parametric models 
of neutron-capture processes will be a first step in establishing if this is a possibility.

If the Ba versus Eu correlation in CEMP-$s/r$ stars is confirmed and if 
it is not due to a new ``$s/r$'' neutron-capture process, but to the $s$ 
process and the $r$ processes occurring in AGB stars and SNIIe,
respectively, then these sites must be somehow
correlated to each other.
A probably unlikely scenario, which we provide here as an illustrative example, 
is that of a stable triple stellar 
system comprising the CEMP-$s/r$ star observed today, a former AGB star, and a former 
SNII. In this case the orbital parameters such as the distance between the stars, 
which affect the mass transfer, are correlated, probably 
resulting in correlated Ba and Eu abundances.

While the AGB models produce a good match to the observed [Mg/hs] ratios in 
CEMP-$s$ stars, they typically produce too high [Na/hs] and [F/hs] ratios. This may depend on 
the choice of the proton profile leading to the formation of the \iso{13}C 
pocket, the proton-capture reaction rates involved, and the occurrence of extra-mixing processes. 
Investigation of these issues may
result in important constraints on these uncertain stellar and nuclear inputs.

Finally, CS\,30322-023 may carry the signature of 
nucleosynthesis in intermediate-mass AGB stars. 
This would be difficult to reconcile with the suggestion of 
\citet{masseron06} that this star 
is an intrinsic AGB object based on the low log\,$g=-0.3$.
This would make this star an extremely peculiar object born 
less than a half a Gyr ago, 
but with an extremely low metallicity typical of the early Universe. 
Our conclusions are 
more in line with the higher value of 
log\,$g=1$ reported by \citet{aoki07}, not indicative of an intrinsic AGB star.
From the preliminary comparison made here the 
composition of CS\,30322-023 indicates that both HBB and the \iso{13}C 
pocket were operating in low-metallicity massive AGB stars. 
However, before drawing 
conclusions we need to compare the composition of this star with predictions from a 
model of the appropriate metallicity of [Fe/H]=$-$3.3.
Observations of Eu in 
CS\,22960-053 and CS\,29528-041 may reveal that these two stars are also  
CEMP-low-$s$ with [N/C]$>$1 creating a new class together with CS\,30322-023.
Future studies will focus on comparison of single stellar objects with our AGB 
model predictions.

\acknowledgments

We heartly thank Sara Bisterzo for lengthy and thorough discussion and extensive criticism, 
which have led to major improvements of this paper and for help in setting up Table 6.  
We thank Thomas Masseron for help on the observational data points and Daniel
Price for discussion on star formation.
We also thank the anonymous referee for helping to improve the discussion
and clarity of the presented results.
ML and AIK are grateful for the support of the NCI National Facility at 
the ANU. 
ML is an ARC Future Fellow and Monash Research Fellow. AIK is a Stromlo Fellow. 
RJS is a Stromlo Fellow and acknowledges support from the Australian Research Council's 
Discovery Projects funding scheme (project number DP0879472) during his time at Monash 
University.

\bibliographystyle{apj}
\bibliography{apj-jour,library}

\begin{thebibliography}{98}
\expandafter\ifx\csname natexlab\endcsname\relax\def\natexlab#1{#1}\fi

\bibitem[{{Abia} {et~al.}(2001){Abia}, {Busso}, {Gallino}, {Dom{\'{\i}}nguez},
  {Straniero}, \& {Isern}}]{abia01}
{Abia}, C., {Busso}, M., {Gallino}, R., {Dom{\'{\i}}nguez}, I., {Straniero},
  O., \& {Isern}, J. 2001, \apj, 559, 1117

\bibitem[{{Angulo} {et~al.}(1999){Angulo}, {Arnould}, {Rayet}, {Descouvemont},
  {Baye}, {Leclercq-Willain}, {Coc}, {Barhoumi}, {Aguer}, {Rolfs}, {Kunz},
  {Hammer}, {Mayer}, {Paradellis}, {Kossionides}, {Chronidou}, {Spyrou},
  {degl'Innocenti}, {Fiorentini}, {Ricci}, {Zavatarelli}, {Providencia},
  {Wolters}, {Soares}, {Grama}, {Rahighi}, {Shotter}, \& {Lamehi
  Rachti}}]{angulo99}
{Angulo}, C., {Arnould}, M., {Rayet}, M., {Descouvemont}, P., {Baye}, D.,
  {Leclercq-Willain}, C., {Coc}, A., {Barhoumi}, S., {Aguer}, P., {Rolfs}, C.,
  {Kunz}, R., {Hammer}, J.~W., {Mayer}, A., {Paradellis}, T., {Kossionides},
  S., {Chronidou}, C., {Spyrou}, K., {degl'Innocenti}, S., {Fiorentini}, G.,
  {Ricci}, B., {Zavatarelli}, S., {Providencia}, C., {Wolters}, H., {Soares},
  J., {Grama}, C., {Rahighi}, J., {Shotter}, A., \& {Lamehi Rachti}, M. 1999,
  \nphysa, 656, 3

\bibitem[{{Aoki} {et~al.}(2007){Aoki}, {Beers}, {Christlieb}, {Norris}, {Ryan},
  \& {Tsangarides}}]{aoki07}
{Aoki}, W., {Beers}, T.~C., {Christlieb}, N., {Norris}, J.~E., {Ryan}, S.~G.,
  \& {Tsangarides}, S. 2007, \apj, 655, 492

\bibitem[{{Arlandini} {et~al.}(1999){Arlandini}, {K{\"a}ppeler}, {Wisshak},
  {Gallino}, {Lugaro}, {Busso}, \& {Straniero}}]{arlandini99}
{Arlandini}, C., {K{\"a}ppeler}, F., {Wisshak}, K., {Gallino}, R., {Lugaro},
  M., {Busso}, M., \& {Straniero}, O. 1999, \apj, 525, 886

\bibitem[{{Asplund} {et~al.}(2009){Asplund}, {Grevesse}, {Sauval}, \&
  {Scott}}]{asplund09}
{Asplund}, M., {Grevesse}, N., {Sauval}, A.~J., \& {Scott}, P. 2009, \araa, 47,
  481

\bibitem[{{Bate}(2009)}]{bate09}
{Bate}, M.~R. 2009, \mnras, 392, 590

\bibitem[{{Beers} \& {Christlieb}(2005)}]{beers05}
{Beers}, T.~C. \& {Christlieb}, N. 2005, \araa, 43, 531

\bibitem[{{Bemmerer} {et~al.}(2006){Bemmerer}, {Confortola}, {Lemut},
  {Bonetti}, {Broggini}, {Corvisiero}, {Costantini}, {Cruz}, {Formicola},
  {F{\"u}l{\"o}p}, {Gervino}, {Guglielmetti}, \& {Gustavino}}]{bemmerer06}
{Bemmerer}, D., {Confortola}, F., {Lemut}, A., {Bonetti}, R., {Broggini}, C.,
  {Corvisiero}, P., {Costantini}, H., {Cruz}, J., {Formicola}, A.,
  {F{\"u}l{\"o}p}, Z., {Gervino}, G., {Guglielmetti}, A., \& {Gustavino}, C.
  2006, Nuclear Physics A, 779, 297

\bibitem[{{Bessell} {et~al.}(1989){Bessell}, {Brett}, {Wood}, \&
  {Scholz}}]{bessell89}
{Bessell}, M.~S., {Brett}, J.~M., {Wood}, P.~R., \& {Scholz}, M. 1989, \aaps,
  77, 1

\bibitem[{{Bisterzo} {et~al.}(2010){Bisterzo}, {Gallino}, {Straniero},
  {Cristallo}, \& {K{\"a}ppeler}}]{bisterzo10}
{Bisterzo}, S., {Gallino}, R., {Straniero}, O., {Cristallo}, S., \&
  {K{\"a}ppeler}, F. 2010, \mnras, 404, 1529

\bibitem[{{Bisterzo} {et~al.}(2011{\natexlab{a}}){Bisterzo}, {Gallino},
  {Straniero}, {Cristallo}, \& {K{\"a}ppeler}}]{bisterzo11b}
---. 2011{\natexlab{a}}, \mnras, submitted

\bibitem[{{Bisterzo} {et~al.}(2011{\natexlab{b}}){Bisterzo}, {Gallino},
  {Straniero}, {Cristallo}, \& {K{\"a}ppeler}}]{bisterzo11a}
---. 2011{\natexlab{b}}, \mnras\ arXiv:1108.0500, 1716

\bibitem[{{Bona{\v c}i{\'c} Marinovi{\'c}} {et~al.}(2007){Bona{\v c}i{\'c}
  Marinovi{\'c}}, {Izzard}, {Lugaro}, \& {Pols}}]{bonacic07}
{Bona{\v c}i{\'c} Marinovi{\'c}}, A., {Izzard}, R.~G., {Lugaro}, M., \& {Pols},
  O.~R. 2007, \aap, 469, 1013

\bibitem[{{Boothroyd} {et~al.}(1993){Boothroyd}, {Sackmann}, \&
  {Ahern}}]{boothroyd93}
{Boothroyd}, A.~I., {Sackmann}, I.-J., \& {Ahern}, S.~C. 1993, \apj, 416, 762

\bibitem[{{Busso} {et~al.}(2001){Busso}, {Gallino}, {Lambert}, {Travaglio}, \&
  {Smith}}]{busso01}
{Busso}, M., {Gallino}, R., {Lambert}, D.~L., {Travaglio}, C., \& {Smith},
  V.~V. 2001, \apj, 557, 802

\bibitem[{{Busso} {et~al.}(1999){Busso}, {Gallino}, \& {Wasserburg}}]{busso99}
{Busso}, M., {Gallino}, R., \& {Wasserburg}, G.~J. 1999, \araa, 37, 239

\bibitem[{{Cameron}(1955)}]{cameron55}
{Cameron}, A.~G.~W. 1955, \apj, 121, 144

\bibitem[{{Campbell} \& {Lattanzio}(2008)}]{campbell08}
{Campbell}, S.~W. \& {Lattanzio}, J.~C. 2008, \aap, 490, 769

\bibitem[{{Cannon}(1993)}]{cannon93}
{Cannon}, R.~C. 1993, \mnras, 263, 817

\bibitem[{{Carollo} {et~al.}(2011){Carollo}, {Beers}, {Bovy}, {Sivarani},
  {Norris}, {Freeman}, {Aoki}, {Lee}, \& {Kennedy}}]{carollo11}
{Carollo}, D., {Beers}, T.~C., {Bovy}, J., {Sivarani}, T., {Norris}, J.~E.,
  {Freeman}, K.~C., {Aoki}, W., {Lee}, Y.~S., \& {Kennedy}, C.~R. 2011, \apj\
  arXiv: 1103.3067

\bibitem[{{Carollo} {et~al.}(2007){Carollo}, {Beers}, {Lee}, {Chiba}, {Norris},
  {Wilhelm}, {Sivarani}, {Marsteller}, {Munn}, {Bailer-Jones}, {Fiorentin}, \&
  {York}}]{carollo07}
{Carollo}, D., {Beers}, T.~C., {Lee}, Y.~S., {Chiba}, M., {Norris}, J.~E.,
  {Wilhelm}, R., {Sivarani}, T., {Marsteller}, B., {Munn}, J.~A.,
  {Bailer-Jones}, C.~A.~L., {Fiorentin}, P.~R., \& {York}, D.~G. 2007, \nat,
  450, 1020

\bibitem[{{Chiosi} {et~al.}(1993){Chiosi}, {Wood}, \& {Capitanio}}]{chiosi93}
{Chiosi}, C., {Wood}, P.~R., \& {Capitanio}, N. 1993, \apjs, 86, 541

\bibitem[{{Clayton}(1988)}]{clayton88}
{Clayton}, D.~D. 1988, \mnras, 234, 1

\bibitem[{{Cohen} {et~al.}(2006){Cohen}, {McWilliam}, {Shectman}, {Thompson},
  {Christlieb}, {Melendez}, {Ramirez}, {Swensson}, \& {Zickgraf}}]{cohen06}
{Cohen}, J.~G., {McWilliam}, A., {Shectman}, S., {Thompson}, I., {Christlieb},
  N., {Melendez}, J., {Ramirez}, S., {Swensson}, A., \& {Zickgraf}, F.-J. 2006,
  \aj, 132, 137

\bibitem[{{Cohen} {et~al.}(2005){Cohen}, {Shectman}, {Thompson}, {McWilliam},
  {Christlieb}, {Melendez}, {Zickgraf}, {Ram{\'{\i}}rez}, \&
  {Swenson}}]{cohen05}
{Cohen}, J.~G., {Shectman}, S., {Thompson}, I., {McWilliam}, A., {Christlieb},
  N., {Melendez}, J., {Zickgraf}, F., {Ram{\'{\i}}rez}, S., \& {Swenson}, A.
  2005, \apjl, 633, L109

\bibitem[{{Cristallo} {et~al.}(2009{\natexlab{a}}){Cristallo}, {Piersanti},
  {Straniero}, {Gallino}, {Dom{\'{\i}}nguez}, \& {K{\"a}ppeler}}]{cristallo09b}
{Cristallo}, S., {Piersanti}, L., {Straniero}, O., {Gallino}, R.,
  {Dom{\'{\i}}nguez}, I., \& {K{\"a}ppeler}, F. 2009{\natexlab{a}}, \pasa, 26,
  139

\bibitem[{{Cristallo} {et~al.}(2009{\natexlab{b}}){Cristallo}, {Straniero},
  {Gallino}, {Piersanti}, {Dom{\'{\i}}nguez}, \& {Lederer}}]{cristallo09a}
{Cristallo}, S., {Straniero}, O., {Gallino}, R., {Piersanti}, L.,
  {Dom{\'{\i}}nguez}, I., \& {Lederer}, M.~T. 2009{\natexlab{b}}, \apj, 696,
  797

\bibitem[{{Denissenkov} \& {Tout}(2003)}]{denissenkov03a}
{Denissenkov}, P.~A. \& {Tout}, C.~A. 2003, \mnras, 340, 722

\bibitem[{{Eggleton}(1971)}]{eggleton71}
{Eggleton}, P.~P. 1971, \mnras, 151, 351

\bibitem[{{Eggleton}(1972)}]{eggleton72}
---. 1972, \mnras, 156, 361

\bibitem[{{Eldridge} \& {Tout}(2004)}]{eldridgetout04}
{Eldridge}, J.~J. \& {Tout}, C.~A. 2004, \mnras, 348, 201

\bibitem[{{Elmegreen}(2011)}]{elmegreen11}
{Elmegreen}, B.~G. 2011, Ecole Evry Schatzman 2010: Star Formation in the Local
  Universe, arXiv:1101.3112

\bibitem[{Ferguson {et~al.}(2005)Ferguson, Alexander, Allard, Barman, Bodnarik,
  Hauschildt, Heffner-Wong, \& Tamanai}]{ferguson05}
Ferguson, J.~W., Alexander, D.~R., Allard, F., Barman, T., Bodnarik, J.~G.,
  Hauschildt, P.~H., Heffner-Wong, A., \& Tamanai, A. 2005, ApJ, 623, 585

\bibitem[{{Frost} \& {Lattanzio}(1996)}]{frost96}
{Frost}, C.~A. \& {Lattanzio}, J.~C. 1996, \apj, 473, 383

\bibitem[{{Gallino} {et~al.}(1998){Gallino}, {Arlandini}, {Busso}, {Lugaro},
  {Travaglio}, {Straniero}, {Chieffi}, \& {Limongi}}]{gallino98}
{Gallino}, R., {Arlandini}, C., {Busso}, M., {Lugaro}, M., {Travaglio}, C.,
  {Straniero}, O., {Chieffi}, A., \& {Limongi}, M. 1998, \apj, 497, 388

\bibitem[{{Goriely} \& {Mowlavi}(2000)}]{goriely00}
{Goriely}, S. \& {Mowlavi}, N. 2000, \aap, 362, 599

\bibitem[{{Goriely} \& {Siess}(2004)}]{goriely04}
{Goriely}, S. \& {Siess}, L. 2004, \aap, 421, L25

\bibitem[{{Goriely} \& {Siess}(2005)}]{goriely05}
{Goriely}, S. \& {Siess}, L. 2005, in IAU Symposium, Vol. 228, From Lithium to
  Uranium: Elemental Tracers of Early Cosmic Evolution, ed. {V.~Hill,
  P.~Fran{\c c}ois, \& F.~Primas}, 451--460

\bibitem[{{Heil} {et~al.}(2008){Heil}, {Detwiler}, {Azuma}, {Couture}, {Daly},
  {G{\"o}rres}, {K{\"a}ppeler}, {Reifarth}, {Tischhauser}, {Ugalde}, \&
  {Wiescher}}]{heil08}
{Heil}, M., {Detwiler}, R., {Azuma}, R.~E., {Couture}, A., {Daly}, J.,
  {G{\"o}rres}, J., {K{\"a}ppeler}, F., {Reifarth}, R., {Tischhauser}, P.,
  {Ugalde}, C., \& {Wiescher}, M. 2008, \prc, 78, 025803

\bibitem[{{Herwig}(2000)}]{herwig00}
{Herwig}, F. 2000, \aap, 360, 952

\bibitem[{{Herwig}(2004{\natexlab{a}})}]{herwig04a}
---. 2004{\natexlab{a}}, \apj, 605, 425

\bibitem[{{Herwig}(2004{\natexlab{b}})}]{herwig04b}
---. 2004{\natexlab{b}}, \apjs, 155, 651

\bibitem[{{Herwig}(2005)}]{herwig05}
---. 2005, \araa, 43, 435

\bibitem[{{Hollowell} \& {Iben}(1988)}]{hollowell88}
{Hollowell}, D. \& {Iben}, Jr., I. 1988, \apjl, 333, L25

\bibitem[{{Iglesias} \& {Rogers}(1996)}]{iglesias96}
{Iglesias}, C.~A. \& {Rogers}, F.~J. 1996, \apj, 464, 943

\bibitem[{{Iliadis} {et~al.}(2010){Iliadis}, {Longland}, {Champagne}, \&
  {Coc}}]{iliadis10}
{Iliadis}, C., {Longland}, R., {Champagne}, A.~E., \& {Coc}, A. 2010, Nuclear
  Physics A, 841, 323

\bibitem[{{Iwamoto} {et~al.}(2004){Iwamoto}, {Kajino}, {Mathews}, {Fujimoto},
  \& {Aoki}}]{iwamoto04}
{Iwamoto}, N., {Kajino}, T., {Mathews}, G.~J., {Fujimoto}, M.~Y., \& {Aoki}, W.
  2004, \apj, 602, 377

\bibitem[{{Jonsell} {et~al.}(2006){Jonsell}, {Barklem}, {Gustafsson},
  {Christlieb}, {Hill}, {Beers}, \& {Holmberg}}]{jonsell06}
{Jonsell}, K., {Barklem}, P.~S., {Gustafsson}, B., {Christlieb}, N., {Hill},
  V., {Beers}, T.~C., \& {Holmberg}, J. 2006, \aap, 451, 651

\bibitem[{{Karakas} \& {Lattanzio}(2007)}]{karakas07b}
{Karakas}, A. \& {Lattanzio}, J.~C. 2007, PASA, 24, 103

\bibitem[{{Karakas}(2010)}]{karakas10a}
{Karakas}, A.~I. 2010, \mnras, 403, 1413

\bibitem[{{Karakas} \& {Lattanzio}(2003)}]{karakas03b}
{Karakas}, A.~I. \& {Lattanzio}, J.~C. 2003, Publ. Astron. Soc. Aust., 20, 279

\bibitem[{{Karakas} {et~al.}(2002){Karakas}, {Lattanzio}, \&
  {Pols}}]{karakas02}
{Karakas}, A.~I., {Lattanzio}, J.~C., \& {Pols}, O.~R. 2002, Publ. Astron. Soc.
  Aust., 19, 515

\bibitem[{{Karakas} {et~al.}(2006){Karakas}, {Lugaro}, {Wiescher}, {Goerres},
  \& {Ugalde}}]{karakas06a}
{Karakas}, A.~I., {Lugaro}, M., {Wiescher}, M., {Goerres}, J., \& {Ugalde}, C.
  2006, \apj, 643, 471

\bibitem[{{Kobayashi} {et~al.}(2011){Kobayashi}, {Karakas}, \&
  {Umeda}}]{kobayashi11}
{Kobayashi}, C., {Karakas}, A.~I., \& {Umeda}, H. 2011, \mnras, 414, 3231

\bibitem[{{Lattanzio}(1986)}]{lattanzio86}
{Lattanzio}, J.~C. 1986, \apj, 311, 708

\bibitem[{{Lattanzio}(1991)}]{lattanzio91}
---. 1991, \apjs, 76, 215

\bibitem[{{Lau} {et~al.}(2009){Lau}, {Stancliffe}, \& {Tout}}]{lau09}
{Lau}, H.~H.~B., {Stancliffe}, R.~J., \& {Tout}, C.~A. 2009, \mnras, 396, 1046

\bibitem[{{Le{\~a}o} {et~al.}(2009){Le{\~a}o}, {de Gouveia Dal Pino},
  {Falceta-Gon{\c c}alves}, {Melioli}, \& {Geraissate}}]{leao09}
{Le{\~a}o}, M.~R.~M., {de Gouveia Dal Pino}, E.~M., {Falceta-Gon{\c c}alves},
  D., {Melioli}, C., \& {Geraissate}, F.~G. 2009, \mnras, 394, 157

\bibitem[{{Lederer} \& {Aringer}(2009)}]{lederer09}
{Lederer}, M.~T. \& {Aringer}, B. 2009, \aap, 494, 403

\bibitem[{{Lucatello} {et~al.}(2006){Lucatello}, {Beers}, {Christlieb},
  {Barklem}, {Rossi}, {Marsteller}, {Sivarani}, \& {Lee}}]{lucatello06}
{Lucatello}, S., {Beers}, T.~C., {Christlieb}, N., {Barklem}, P.~S., {Rossi},
  S., {Marsteller}, B., {Sivarani}, T., \& {Lee}, Y.~S. 2006, \apjl, 652, L37

\bibitem[{{Lucatello} {et~al.}(2011){Lucatello}, {Masseron}, {Johnson},
  {Pignatari}, \& {Herwig}}]{lucatello11}
{Lucatello}, S., {Masseron}, T., {Johnson}, J.~A., {Pignatari}, M., \&
  {Herwig}, F. 2011, \apj, 729, 40

\bibitem[{{Lucatello} {et~al.}(2005){Lucatello}, {Tsangarides}, {Beers},
  {Carretta}, {Gratton}, \& {Ryan}}]{lucatello05b}
{Lucatello}, S., {Tsangarides}, S., {Beers}, T.~C., {Carretta}, E., {Gratton},
  R.~G., \& {Ryan}, S.~G. 2005, \apj, 625, 825

\bibitem[{{Lugaro} {et~al.}(2009){Lugaro}, {Campbell}, \& {de Mink}}]{lugaro09}
{Lugaro}, M., {Campbell}, S.~W., \& {de Mink}, S.~E. 2009, \pasa, 26, 322

\bibitem[{{Lugaro} {et~al.}(2008){Lugaro}, {de Mink}, {Izzard}, {Campbell},
  {Karakas}, {Cristallo}, {Pols}, {Lattanzio}, {Straniero}, {Gallino}, \&
  {Beers}}]{lugaro08a}
{Lugaro}, M., {de Mink}, S.~E., {Izzard}, R.~G., {Campbell}, S.~W., {Karakas},
  A.~I., {Cristallo}, S., {Pols}, O.~R., {Lattanzio}, J.~C., {Straniero}, O.,
  {Gallino}, R., \& {Beers}, T.~C. 2008, \aap, 484, L27

\bibitem[{{Lugaro} {et~al.}(2003){Lugaro}, {Herwig}, {Lattanzio}, {Gallino}, \&
  {Straniero}}]{lugaro03b}
{Lugaro}, M., {Herwig}, F., {Lattanzio}, J.~C., {Gallino}, R., \& {Straniero},
  O. 2003, \apj, 586, 1305

\bibitem[{{Lugaro} {et~al.}(2004){Lugaro}, {Ugalde}, {Karakas}, {G{\"o}rres},
  {Wiescher}, {Lattanzio}, \& {Cannon}}]{lugaro04}
{Lugaro}, M., {Ugalde}, C., {Karakas}, A.~I., {G{\"o}rres}, J., {Wiescher}, M.,
  {Lattanzio}, J.~C., \& {Cannon}, R.~C. 2004, \apj, 615, 934

\bibitem[{{Marigo}(2002)}]{marigo02}
{Marigo}, P. 2002, \aap, 387, 507

\bibitem[{{Masseron} {et~al.}(2010){Masseron}, {Johnson}, {Plez}, {van Eck},
  {Primas}, {Goriely}, \& {Jorissen}}]{masseron10}
{Masseron}, T., {Johnson}, J.~A., {Plez}, B., {van Eck}, S., {Primas}, F.,
  {Goriely}, S., \& {Jorissen}, A. 2010, \aap, 509, A93+

\bibitem[{{Masseron} {et~al.}(2006){Masseron}, {van Eck}, {Famaey}, {Goriely},
  {Plez}, {Siess}, {Beers}, {Primas}, \& {Jorissen}}]{masseron06}
{Masseron}, T., {van Eck}, S., {Famaey}, B., {Goriely}, S., {Plez}, B.,
  {Siess}, L., {Beers}, T.~C., {Primas}, F., \& {Jorissen}, A. 2006, \aap, 455,
  1059

\bibitem[{{Merrill}(1952)}]{merrill52}
{Merrill}, S.~P.~W. 1952, \apj, 116, 21

\bibitem[{{Montes} {et~al.}(2007){Montes}, {Beers}, {Cowan}, {Elliot},
  {Farouqi}, {Gallino}, {Heil}, {Kratz}, {Pfeiffer}, {Pignatari}, \&
  {Schatz}}]{montes07}
{Montes}, F., {Beers}, T.~C., {Cowan}, J., {Elliot}, T., {Farouqi}, K.,
  {Gallino}, R., {Heil}, M., {Kratz}, K., {Pfeiffer}, B., {Pignatari}, M., \&
  {Schatz}, H. 2007, \apj, 671, 1685

\bibitem[{{Otsuka} {et~al.}(2010){Otsuka}, {Tajitsu}, {Hyung}, \&
  {Izumiura}}]{otsuka10}
{Otsuka}, M., {Tajitsu}, A., {Hyung}, S., \& {Izumiura}, H. 2010, \apj, 723,
  658

\bibitem[{{Pignatari} {et~al.}(2008){Pignatari}, {Gallino}, {Meynet},
  {Hirschi}, {Herwig}, \& {Wiescher}}]{pignatari08}
{Pignatari}, M., {Gallino}, R., {Meynet}, G., {Hirschi}, R., {Herwig}, F., \&
  {Wiescher}, M. 2008, \apjl, 687, L95

\bibitem[{{Pols} {et~al.}(1995){Pols}, {Tout}, {Eggleton}, \& {Han}}]{pols95}
{Pols}, O.~R., {Tout}, C.~A., {Eggleton}, P.~P., \& {Han}, Z. 1995, MNRAS, 274,
  964

\bibitem[{{Reimers}(1975)}]{reimers75}
{Reimers}, D. 1975, {Circumstellar envelopes and mass loss of red giant stars}
  (Problems in stellar atmospheres and envelopes.), 229--256

\bibitem[{{Roberts} {et~al.}(2010){Roberts}, {Woosley}, \&
  {Hoffman}}]{roberts10}
{Roberts}, L.~F., {Woosley}, S.~E., \& {Hoffman}, R.~D. 2010, \apj, 722, 954

\bibitem[{{Ryan} {et~al.}(2005){Ryan}, {Aoki}, {Norris}, \& {Beers}}]{ryan05}
{Ryan}, S.~G., {Aoki}, W., {Norris}, J.~E., \& {Beers}, T.~C. 2005, \apj, 635,
  349

\bibitem[{{Smith} \& {Lambert}(1986)}]{smith86b}
{Smith}, V.~V. \& {Lambert}, D.~L. 1986, \apj, 311, 843

\bibitem[{{Sneden} {et~al.}(2008){Sneden}, {Cowan}, \& {Gallino}}]{sneden08}
{Sneden}, C., {Cowan}, J.~J., \& {Gallino}, R. 2008, \araa, 46, 241

\bibitem[{{Sneden} {et~al.}(2003){Sneden}, {Cowan}, {Lawler}, {Ivans},
  {Burles}, {Beers}, {Primas}, {Hill}, {Truran}, {Fuller}, {Pfeiffer}, \&
  {Kratz}}]{sneden03}
{Sneden}, C., {Cowan}, J.~J., {Lawler}, J.~E., {Ivans}, I.~I., {Burles}, S.,
  {Beers}, T.~C., {Primas}, F., {Hill}, V., {Truran}, J.~W., {Fuller}, G.~M.,
  {Pfeiffer}, B., \& {Kratz}, K. 2003, \apj, 591, 936

\bibitem[{{Stancliffe}(2009)}]{stancliffe09}
{Stancliffe}, R.~J. 2009, \mnras, 394, 1051

\bibitem[{{Stancliffe}(2010)}]{stancliffe10}
---. 2010, \mnras, 403, 505

\bibitem[{{Stancliffe} {et~al.}(2011){Stancliffe}, {Dearborn}, {Lattanzio},
  {Heap}, \& {Campbell}}]{stancliffe11}
{Stancliffe}, R.~J., {Dearborn}, D.~S.~P., {Lattanzio}, J.~C., {Heap}, S.~A.,
  \& {Campbell}, S.~W. 2011, \apj\ arXiv: 1109.1289

\bibitem[{{Stancliffe} \& {Eldridge}(2009)}]{stancliffeeldridge09}
{Stancliffe}, R.~J. \& {Eldridge}, J.~J. 2009, \mnras, 396, 1699

\bibitem[{{Stancliffe} \& {Glebbeek}(2008)}]{stancliffeglebbeek08}
{Stancliffe}, R.~J. \& {Glebbeek}, E. 2008, \mnras, 389, 1828

\bibitem[{{Stancliffe} {et~al.}(2007){Stancliffe}, {Glebbeek}, {Izzard}, \&
  {Pols}}]{stancliffe07a}
{Stancliffe}, R.~J., {Glebbeek}, E., {Izzard}, R.~G., \& {Pols}, O.~R. 2007,
  \aap, 464, L57

\bibitem[{{Stancliffe} {et~al.}(2005){Stancliffe}, {Lugaro}, {Ugalde}, {Tout},
  {G{\"o}rres}, \& {Wiescher}}]{stancliffe05b}
{Stancliffe}, R.~J., {Lugaro}, M., {Ugalde}, C., {Tout}, C.~A., {G{\"o}rres},
  J., \& {Wiescher}, M. 2005, \mnras, 360, 375

\bibitem[{{Stancliffe} {et~al.}(2004){Stancliffe}, {Tout}, \&
  {Pols}}]{stancliffe04b}
{Stancliffe}, R.~J., {Tout}, C.~A., \& {Pols}, O.~R. 2004, \mnras, 352, 984

\bibitem[{{Straniero} {et~al.}(1995){Straniero}, {Gallino}, {Busso}, {Chiefei},
  {Raiteri}, {Limongi}, \& {Salaris}}]{straniero95}
{Straniero}, O., {Gallino}, R., {Busso}, M., {Chiefei}, A., {Raiteri}, C.~M.,
  {Limongi}, M., \& {Salaris}, M. 1995, \apjl, 440, L85

\bibitem[{{Suda} \& {Fujimoto}(2010)}]{suda10}
{Suda}, T. \& {Fujimoto}, M.~Y. 2010, \mnras, 405, 177

\bibitem[{{Suda} {et~al.}(2008){Suda}, {Katsuta}, {Yamada}, {Suwa}, {Ishizuka},
  {Komiya}, {Sorai}, {Aikawa}, \& {Fujimoto}}]{suda08}
{Suda}, T., {Katsuta}, Y., {Yamada}, S., {Suwa}, T., {Ishizuka}, C., {Komiya},
  Y., {Sorai}, K., {Aikawa}, M., \& {Fujimoto}, M.~Y. 2008, \pasj, 60, 1159

\bibitem[{{Travaglio} {et~al.}(2004){Travaglio}, {Gallino}, {Arnone}, {Cowan},
  {Jordan}, \& {Sneden}}]{travaglio04}
{Travaglio}, C., {Gallino}, R., {Arnone}, E., {Cowan}, J., {Jordan}, F., \&
  {Sneden}, C. 2004, \apj, 601, 864

\bibitem[{{Truran} \& {Iben}(1977)}]{iben77}
{Truran}, J.~W. \& {Iben}, Jr., I. 1977, \apj, 216, 797

\bibitem[{{Van Eck} {et~al.}(2003){Van Eck}, {Goriely}, {Jorissen}, \&
  {Plez}}]{vaneck03}
{Van Eck}, S., {Goriely}, S., {Jorissen}, A., \& {Plez}, B. 2003, \aap, 404,
  291

\bibitem[{{van Raai} {et~al.}(2011){van Raai}, {Lugaro}, {Karakas},
  {Garc{\'{\i}}a-Hern{\'a}ndez}, \& {Yong}}]{vanraai11}
{van Raai}, M., {Lugaro}, M., {Karakas}, A.~I., {Garc{\'{\i}}a-Hern{\'a}ndez},
  D.~A., \& {Yong}, D. 2011, \aap, submitted

\bibitem[{{Vanhala} \& {Cameron}(1998)}]{vanhala98}
{Vanhala}, H.~A.~T. \& {Cameron}, A.~G.~W. 1998, \apj, 508, 291

\bibitem[{{Vassiliadis} \& {Wood}(1993)}]{vw93}
{Vassiliadis}, E. \& {Wood}, P.~R. 1993, \apj, 413, 641

\bibitem[{{Wallerstein} \& {Knapp}(1998)}]{wallerstein98}
{Wallerstein}, G. \& {Knapp}, G.~R. 1998, \araa, 36, 369

\end{thebibliography}

%%\clearpage

%%\clearpage

%% If you are not including electonic art with your submission, you may
%% mark up your captions using the \figcaption command. See the
%% No more than seven \figcaption commands are allowed per page,
%% so if you have more than seven captions, insert a \clearpage

%% Tables should be submitted one per page, so put a \clearpage before

%%\clearpage

%% Two options are available to the author for producing tables:  the
%% deluxetable environment provided by the AASTeX package or the LaTeX
%% table environment.  Use of deluxetable is preferred.
%%

%% Three table samples follow, two marked up in the deluxetable environment,
%% one marked up as a LaTeX table.

%% In this first example, note that the \tabletypesize{}
%% command has been used to reduce the font size of the table.
%% We also use the \rotate command to rotate the table to
%% landscape orientation since it is very wide even at the
%% Note also that the \label command needs to be placed

%% This table also includes a table comment indicating that the full
%% version will be available in machine-readable format in the electronic

%\newpage

\begin{center}
\begin{sidewaystable}
\caption{Details of stellar models computed with the {\sc stars} code.}
\label{tab:starscode}
   \begin{tabular}{ c c c c c c c c c }
     \hline
Mass (\msun) & TPs & TPs with TDU & T$^{\rm max}_{\rm He}$ (MK) & $\lambda_{\rm max}$ &
total M$_{\rm dred}$ (\msun) & final M$_{\rm env}$ (\msun) & HBB
& $T_{\rm max}^{\rm bce}$ (MK) \\
\hline
1.0 & 13 & 3-8 & 273 & 0.127 & 0.006 & 0.016 & No & 1.64 \\
1.25 & 9 & 2-7 & 267 & 0.431 & 0.021 & 0.015 & No & 2.75 \\
1.5 & 9 & 2-8 & 268 & 0.676 & 0.040 & 0.408 & No & 3.97 \\
1.75 & 11 & 2-10 & 276 & 0.846 & 0.074 & 0.08 & No & 6.05 \\
2.0 & 13 & 2-12 & 283 & 0.971 & 0.098 & 0.468 & No & 9.79 \\
2.25 & 13 & 2-13 & 294 & 1.060 & 0.082 & 0.584 & No & 22.04 \\
2.5 & 16 & 2-16 & 298 & 1.083 & 0.107 & 0.789 & No & 32.05 \\
3.0 & 16 & 2-16 & 331 & 1.116 & 0.047 & 0.979 & Yes & 56.76 \\
\hline
   \end{tabular}
   \end{sidewaystable}
\end{center}

\begin{center}
\begin{sidewaystable}
\caption{Details of stellar models computed with the Stromlo code.}
\label{tab:stromlocode}
  \begin{tabular}{ c c c c c c c c c }
    \hline
Mass (\msun) & TPs & TPs with TDU & T$^{\rm max}_{\rm He}$ (MK) & $\lambda_{\rm max}$ &
total M$_{\rm dred}$ (\msun) & final M$_{\rm env}$ (\msun) & HBB
& $T_{\rm max}^{\rm bce}$ (MK) \\
\hline
0.9 & 38 & 4-20 & 278$^a$ & 0.14 & 0.012 & 0.029 & No & 1.8  \\
1.0 & 26 & 5-6 & 261$^b$ & 0.18 & 0.003 & 0.0085 & No & 1.5 \\
1.25 & 16 & 3-13 & 274$^c$ & 0.25 & 0.020 & 0.018 & No & 2.3 \\
1.5 & 18 & 3-17 & 282 & 0.51 & 0.059 & 0.022 & No & 3.4 \\
1.75 & 20 & 3-20 & 294 & 0.71 & 0.132 & 0.029 & No & 5.4 \\
1.9 & 24 & 3-23 & 300 & 0.80 & 0.169 & 0.029 & No & 7.4 \\
2.0 & 26 & 2-26 & 307 & 0.85 & 0.218 & 0.038 & No & 9.0 \\ % old model
  & 19$^d$ & 2-18 & 305 & 0.85 & 0.133 & 0.0$^e$ & No & 6.6 \\ %updated model
2.25 & 33 & 3-33 & 320 & 0.94 & 0.292 & 0.162 & No & 15.7 \\
2.5 & 30 & 2-30 & 351 & 1.01 & 0.283 & 0.413 & No & 28.4 \\ %but the nucleosynthesis goes till Menv ~1.65
    & 21$^d$ & 1-20 & 341 & 0.99 & 0.165 & 0.0$^e$ & No & 16.3 \\ %updated model, Menv ~ 3.6e-5
3.0 & 20 & 2-20 & 336 & 1.02 & 0.106 & 0.119 & Yes & 39.6 \\ % mild HBB only
3.5 & 27 & 2-27 & 348 & 1.04 & 0.103 & 0.136 & Yes & 75.9 \\
4.0 & 37 & 2-37 & 352 & 0.97 & 0.128 & 0.133 & Yes & 82.9  \\
4.5 & 41 & 2-41 & 352 & 0.96 & 0.110 & 0.085 & Yes & 87.2  \\
5.0 & 56 & 2-56 & 363 & 0.96 & 0.106 & 0.124 & Yes & 92.5 \\
5.5 & 77 & 2-77 & 369 & 0.94 & 0.124 & 0.085  & Yes & 98.2 \\
6.0 & 109 & 2-109 & 378 & 0.93 & 0.127 & 0.183 & Yes & 105 \\
\hline
  \end{tabular}
$^a$This temperature refers to the TP followed by TDU. The absolute maximum
temperature was 302~MK.
$^b$This temperature refers to the TP followed by TDU. The absolute maximum
temperature was 298~MK.
$^c$This temperature refers to the TP followed by TDU. The absolute maximum
temperature was 277~MK.
$^d$This model was computed using the C and N-rich opacity tables
from \citet{lederer09}. 
$^e$This model was evolved to the white dwarf cooling track. The final
envelope mass was a few $10^{-5}$ \msun, which is the remaining
H-shell.
\end{sidewaystable}
\end{center}

%\newpage

\begin{center}
\begin{sidewaystable}
\caption{Selected key elemental abundances [X/Fe] at the stellar surface at the 
end of the computed evolution for all our {\sc stars} models 
with a scaled solar initial composition from Asplund et al (2009).}
\label{tab:resultsSTARS}
  \begin{tabular}{ c c c c c c c c c c c }
    \hline
Mass (\msun) & M$_{\rm mix}$ (\msun) & [Na/Fe] & [Mg/Fe] & [Sr/Fe] & [Ba/Fe] & 
[Eu/Fe] & [Pb/Fe] & [Ba/Sr] & [Ba/Eu] & [Pb/Ba] \\
\hline
1.25 & 0 & 0.65 & 0.08 & 0.36 & 0.02 & $-$0.01 & 0.02 & $-$0.34 & 0.03 & 0.00 \\
 & 0.002 & 1.62 & 0.51 & 1.65 & 1.59 & 0.69 & 2.75 & $-$0.06 & 0.90 & 1.16 \\ 
1.5 & 0 & 0.92 & 0.21 & 0.07 & 0.02 & $-$0.01 & 0.01 & $-$0.05 & 0.03 & $-$0.01 \\
  & 0.0006 & 1.31 & 0.42 & 1.46 & 1.38 & 0.47 & 2.49 & $-$0.08 & 0.91 & 1.11 \\ 
 & 0.002 & 1.86 & 0.76 & 1.64 & 1.78 & 0.85 & 2.91 & 0.14 & 0.93 & 1.13 \\ 
 & 0.004 & 2.28 & 1.03 & 1.34 & 2.03 & 1.11 & 3.10 & 0.69 & 0.92 & 1.07 \\ 
1.75 & 0 & 1.58 & 0.68 & 0.10 & 0.04 & $-$0.02 & 0.02 & $-$0.06 & 0.12 & $-$0.02 \\
  & 0.002 & 2.24 & 1.15 & 1.25 & 1.94 & 1.03 & 3.05 & 0.69 & 0.91 & 1.11 \\ 
2.0 & 0 & 1.65 & 0.92 & 0.30 & 0.64 & 0.08 & 0.44 & 0.34 & 0.56 & $-$0.20 \\
 & 0.0006 & 1.90 & 1.09 & 0.90 & 1.52 & 0.67 & 2.69 & 0.62 & 0.85 & 1.17 \\ 
 & 0.002 & 2.26 & 1.32 & 1.34 & 2.04 & 1.12 & 3.03 & 0.70 & 0.92 & 0.99 \\ 
 & 0.004 & 2.53 & 1.52 & 1.59 & 2.31 & 1.38 & 3.12 & 0.72 & 0.93 & 0.81 \\ 
2.25 & 0.002 & 1.86 & 1.38 & 1.18 & 1.85 & 0.92 & 2.72 & 0.67 & 0.93 & 0.87 \\
2.5 & 0.001 & 1.40 & 1.24 & 0.85 & 1.47 & 0.60 & 2.48 & 0.62 & 0.87 & 1.01 \\
 & 0.002 & 1.64 & 1.42 & 1.06 & 1.80 & 0.88 & 2.59 & 0.74 & 0.92 & 0.79 \\ 
3.0 & 0 & 0.67 & 0.69 & 0.47 & 0.11 & 0. & 0.02 & $-$0.36 & 0.11 & $-$0.09 \\ 
 & 0.0005 & 1.23 & 1.19 & 0.83 & 1.25 & 0.42 & 2.24 & 0.42 & 0.83 & 0.99 \\ 
    \hline
  \end{tabular}
  \end{sidewaystable}
\end{center}

\begin{center}
\begin{sidewaystable}
\caption{Selected key elemental abundances [X/Fe] at the stellar surface at the 
end of the computed evolution for all our Stromlo models with a scaled 
solar initial composition from Asplund et al. (2009).}
\label{tab:resultsMM}
  \begin{tabular}{ c c c c c c c c c c c }
    \hline
Mass (\msun) & M$_{\rm mix}$ (\msun) & [Na/Fe] & [Mg/Fe] & [Sr/Fe] & [Ba/Fe] & [Eu/Fe] &
[Pb/Fe] & [Ba/Sr] & [Ba/Eu] & [Pb/Ba] \\
\hline
0.9 & 0 & 1.27 & 0.13 & 1.55 & 2.24 & 1.05 & 1.85 & 0.69 & 1.19 & $-$0.39 \\
 & 0.002 & 2.07 & 0.67 & 1.56 & 2.35 & 1.31 & 3.02 & 0.79 & 1.04 & 0.67 \\
1.0 & 0 & 0.30 & 0.02 & 1.14 & 2.05 & 1.08 & 2.29 & 0.91 & 0.97 & 0.24 \\ 
 & 0.002 & 0.64 & 0.02 & 1.10 & 2.03 & 1.11 & 2.46 & 0.93 & 0.92 & 0.43 \\ 
1.25 & 0 & 0.60 & 0.05 & 1.37 & 0.76 & 0.03 & 0.07 & $-$0.61 & 0.73 & $-$0.69 \\ 
 & 0.002 & 1.68 & 0.50 & 1.51 & 2.17 & 1.08 & 2.76 & 0.66 & 1.09 & 0.59 \\ 
1.5 & 0 & 1.51 & 0.31 & 1.57 & 2.12 & 0.97 & 1.80 & 0.55 & 1.15 & $-$0.32 \\ 
 & 0.0006 & 1.78 & 0.55 & 1.59 & 2.38 & 1.37 & 2.88 & 0.79 & 1.01 & 0.50 \\ 
 & 0.002 & 2.30 & 0.91 & 1.64 & 2.45 & 1.47 & 3.15 & 0.81 & 0.98 & 0.70 \\ 
 & 0.004 & 2.68 & 1.20 & 1.78 & 2.51 & 1.53 & 3.20 & 0.73 & 0.98 & 0.69 \\
1.75 & 0 & 2.30 & 0.85 & 1.83 & 2.27 & 0.96 & 1.55 & 0.44 & 1.31 & $-$0.72 \\
 & 0.002 & 2.75 & 1.28 & 1.93 & 2.54 & 1.46 & 3.22 & 0.61 & 1.08 & 0.68 \\ 
1.9 & 0.002 & 2.90 & 1.45 & 1.90 & 2.42 & 1.45 & 3.23 & 0.52 & 0.97 & 0.81 \\ 
2.0 & 0 & 2.73 & 1.35 & 0.62 & 0.12 & $-$0.05 & 0.04 & $-$0.50 & 0.17 & $-$0.08 \\
 & 0$^a$ & 2.09 & 0.91 & 0.57 & 0.16 & $-$0.03 & 0.03 & $-$0.41 & 0.19 & $-$0.13 \\
 & 0.0006 & 2.81 & 1.45 & 1.51 & 1.89 & 0.98 & 2.92 & 0.38 & 0.91 & 1.03 \\ 
 & 0.002 & 3.02 & 1.62 & 1.78 & 2.42 & 1.51 & 3.24 & 0.64 & 0.91 & 0.82 \\ 
 & 0.002$^a$ & 2.57 & 1.30 & 1.56 & 2.21 & 1.30 & 3.17 & 0.65 & 0.91 & 0.96 \\
 & 0.004 & 3.21 & 1.80 & 1.98 & 2.65 & 1.68 & 3.28 & 0.67 & 0.97 & 0.63 \\ 
2.25 & 0.002 & 3.07 & 1.94 & 1.81 & 2.51 & 1.55 & 3.20 & 0.70 & 0.96 & 0.69 \\ 
2.5 & 0.001 & 2.60 & 2.05 & 1.47 & 2.11 & 1.18 & 3.04 & 0.64 & 0.93 & 0.93 \\ 
 & 0.002 & 2.71 & 2.13 & 1.72 & 2.40 & 1.44 & 3.14 & 0.68 & 0.96 & 0.74 \\ 
 & 0.002$^a$ & 2.28 & 1.67 & 1.50 & 2.21 & 1.26 & 3.04 & 0.71 & 0.95 & 0.83 \\
\hline
  \end{tabular}
$^a$Model computed with low-temperature opacities from \citet{lederer09}.
  \end{sidewaystable}
\end{center}

\setcounter{table}{3}
\begin{center}
\begin{sidewaystable}
\caption{Continues.}
%\vspace{0.5cm}
\label{}
  \begin{tabular}{ c c c c c c c c c c c }
    \hline
Mass (\msun) & M$_{\rm mix}$ (\msun) & [Na/Fe] & [Mg/Fe] & [Sr/Fe] & [Ba/Fe] & [Eu/Fe] &
[Pb/Fe] & [Ba/Sr] & [Ba/Eu] & [Pb/Ba] \\
\hline
3.0 & 0 & 1.91 & 1.72 & 1.40 & 0.79 & 0.04 & 0.19 & $-$0.61 & 0.75 & $-$0.60 \\ 
 & 0.0005 & 2.08 & 1.87 & 1.43 & 1.76 & 0.78 & 2.77 & 0.33 & 0.98 & 1.01 \\ 
 & 0.001 & 2.17 & 1.92 & 1.51 & 2.06 & 1.08 & 2.90 & 0.55 & 0.98 & 0.84 \\ 
3.5 & 0 & 1.75 & 1.59 & 1.45 & 1.01 & 0.09 & 0.34 & $-$0.44 & 0.92 & $-$0.67 \\ 
4.0 & 0 & 1.91 & 1.70 & 1.58 & 1.25 & 0.18 & 0.50 & $-$0.33 & 1.07 & $-$0.75 \\ 
4.5 & 0 & 1.75 & 1.59 & 1.59 & 1.34 & 0.22 & 0.58 & $-$0.25 & 1.12 & $-$0.76 \\ 
5.0 & 0 & 1.78 & 1.74 & 1.67 & 1.35 & 0.20 & 0.63 & $-$0.32 & 1.15 & $-$0.72 \\ 
5.5 & 0 & 1.81 & 1.95 & 1.84 & 1.54 & 0.29 & 0.82 & $-$0.30 & 1.25 & $-$0.72 \\ 
 & 0.0001 & 1.80 & 1.94 & 1.76 & 1.60 & 0.50 & 2.49 & $-$0.16 & 1.10 & 0.89 \\
6.0 & 0 & 1.76 & 1.99 & 1.93 & 1.72 & 0.42 & 0.96 & $-$0.21 & 1.30 & $-$0.76 \\ 
\hline
  \end{tabular}
  \end{sidewaystable}
\end{center}

\begin{sidewaystable}
\caption{Elemental abundances [X/Fe] for C, N, O, F, and Ne at the stellar surface at the 
end of the computed evolution for our models with a scaled solar initial composition 
from Asplund et al. (2009) and M$_{\rm mix}$=0.002 \msun\ for 
Mass $<$ 3 \msun\ and M$_{\rm mix}$=0 for Mass $\geq$ 3 \msun.}
\label{tab:resultslight}
\begin{center}
  \begin{tabular}{ c c c c c c c c c c c }
    \hline
Mass (\msun) & \multicolumn{5}{c}{Stromlo} & \multicolumn{5}{c}{\sc stars} \\
& [C/Fe] & [N/Fe] & [O/Fe] & [F/Fe] & [Ne/Fe] & [C/Fe] & [N/Fe] & [O/Fe] & [F/Fe] & [Ne/Fe] \\
\hline
0.9 & 2.74 & 1.82 & 1.07 & 2.78 & 2.06 \\
1.0 & 1.86 & 0.87 & 0.42 & 1.52 & 0.68 & 2.56 & 1.37 & 1.15 & 2.75 & 1.88 \\
1.25 & 2.60 & 0.98 & 0.91 & 2.53 & 1.84 & 2.75 & 1.06 & 1.22 & 2.82 & 1.90 \\
1.5 & 2.90 & 1.10 & 1.17 & 3.02 & 2.30 & 2.84 & 0.86 & 1.28 & 3.04 & 2.09 \\
1.75 & 3.10 & 1.08 & 1.27 & 3.40 & 2.61 & 2.98 & 0.83 & 1.31 & 3.39 & 2.35 \\
1.9 & 3.15 & 1.14 & 1.29 & 3.54 & 2.72 \\
2.0$^a$ & 3.21 & 1.14 & 1.31 & 3.63 & 2.82 & 2.78 & 0.66 & 1.05 & 3.27 & 2.17 \\
2.0$^b$ & 3.05 & 1.12 & 1.17 & 3.38 & 2.56 & & & & & \\
2.25 & 3.27 & 1.06 & 1.34 & 3.73 & 2.93 & 2.65 & 0.51 & 0.92 & 2.65 & 2.05 \\
2.5 & 3.20 & 0.85 & 1.30 & 3.43 & 2.77 & 2.60 & 0.40 & 0.84 & 3.10 & 1.95 \\
2.5$^b$ & 3.00 & 0.80 & 1.11 & 3.19 & 2.49 & & & & & \\
3.0 & 3.03 & 1.46 & 0.98 & 2.34 & 2.00 & 2.41 & 1.56 & 0.65 & 1.72 & 0.83 \\
3.5 & 2.60 & 3.26 & 0.95 & 1.60 & 1.75 \\
4.0 & 2.69 & 3.43 & 1.29 & 2.21 & 1.95 \\
4.5 & 2.55 & 3.35 & 1.20 & 1.80 & 1.75 \\
5.0 & 2.40 & 3.38 & 0.91 & 0.83 & 1.70 \\
5.5 & 2.46 & 3.42 & 1.00 & 0.09 & 1.67 \\
6.0 & 2.28 & 3.49 & 1.24 & $-$0.45 & 1.65 \\
\hline
  \end{tabular}
\end{center}
%$^a$For comparison, the M=1.3 \msun\ models by \citet{bisterzo10} produce [C/Fe]=3.13, [N/Fe]=0.56 - 
%0.73, [O/Fe]=1.12 - 1.25, and [F/Fe]=2.60 - 3.25.
%$^b$For comparison, the M=1.5 \msun\ models by \citet{bisterzo10} produce [C/Fe]=3.13, [N/Fe]=0.56 -
%0.73, [O/Fe]=1.12 - 1.25, and [F/Fe]=2.60 - 3.25.
$^a$For comparison, the $M$=2 \msun\ model by \citet{cristallo09a} 
produces [C/Fe]=3.00, [N/Fe]=0.86, [O/Fe]=1.01, [F/Fe]=2.90, and [Ne/Fe]=2.27.
$^b$Model computed with low-temperature opacities from \citet{lederer09}.
  \end{sidewaystable}

%\newpage

% AK calling my models Stromlo models, because that is really more appropriate :-)

\begin{center}
\begin{sidewaystable}
\caption{Selected key elemental abundances [X/Fe] at the stellar
surface at the end of the computed evolution for different 
models.
(SM) = Stromlo; (S) = {\sc stars} models; (B10) = \citet{bisterzo10}; (C09) =
\citet{cristallo09a}.}
\label{tab:comparemodels}
   \begin{tabular}{ c c c c c c c c c c c }
     \hline
Mass (\msun) & M$_{\rm mix}$(\msun) & [Na/Fe] & [Mg/Fe] & [Sr/Fe] & [Ba/Fe] &
[Eu/Fe] & [Pb/Fe] & [Ba/Sr] & [Ba/Eu] & [Pb/Ba] \\
\hline
1.25 (SM) & 0.002 & 1.68 & 0.50 & 1.51 & 2.17 & 1.08 & 2.76 & 0.66 & 1.09 & 0.59 \\
1.25 (S) & 0.002 & 1.62 & 0.51 & 1.65 & 1.59 & 0.69 & 2.75 & $-$0.06 & 0.90 & 1.16 \\ 
1.3 (B10) & $\sim$0.001, ST$^a$ & 0.66 & 0.38 & 0.32 & 0.78 & 0.21 & 3.22 & 0.46 & 0.57 & 2.44 \\
          & $\sim$0.001, ST/12 & 0.48 & 0.31 & 0.92 & 2.05 & 1.17 & 2.85 & 1.13 & 0.88 & 0.80 \\
\hline
1.5 (SM) & 0.0006 & 1.78 & 0.55 & 1.59 & 2.38 & 1.37 & 2.88 & 0.79 & 1.01 & 0.50 \\ 
         & 0.002  & 2.30 & 0.91 & 1.64 & 2.45 & 1.47 & 3.15 & 0.81 & 0.98 & 0.70 \\
1.5 (S) & 0.0006  & 1.31 & 0.42 & 1.46 & 1.38 & 0.47 & 2.49 & $-$0.08 & 0.91 & 1.11 \\ 
       & 0.002 & 1.86 & 0.76 & 1.64 & 1.78 & 0.85 & 2.91 & 0.14 & 0.93 & 1.13 \\ 
1.5 (B10) & $\sim$0.001, ST & 2.26 & 1.69 & 1.34 & 2.04 & 1.28 & 4.06 & 0.70 & 0.76 & 2.02 \\
          & $\sim$0.001, ST/12 & 2.27 & 1.47 & 2.26 & 2.73 & 1.71 & 3.11 & 0.47 & 1.02 & 0.38 \\
\hline
2.0 (SM) & 0.0006 & 2.81 & 1.45 & 1.51 & 1.89 & 0.98 & 2.92 & 0.38 & 0.91 & 1.03 \\ 
         & 0.002 & 3.02 & 1.62 & 1.78 & 2.42 & 1.51 & 3.24 & 0.64 & 0.91 & 0.82 \\
         & 0.002$^b$ & 2.46 & 1.24 & 1.72 & 2.46 & 1.40 & 3.16 & 0.74 & 1.06 & 0.70 \\
2.0 (S) & 0.0006 & 1.90 & 1.09 & 0.90 & 1.52 & 0.67 & 2.69 & 0.62 & 0.85 & 1.17 \\ 
        & 0.002 & 2.26 & 1.32 & 1.34 & 2.04 & 1.12 & 3.03 & 0.70 & 0.92 & 0.99 \\ 
2.0 (B10) & $\sim$0.001, ST & 1.89 & 1.44 & 1.34 & 1.87 & 1.05 & 4.12 & 0.53 & 0.82 & 2.25 \\
         & $\sim$0.001, ST/12 & 1.87 & 1.19 & 2.15 & 2.96 & 1.98 & 3.44 & 0.81 & 0.98 & 0.48 \\
2.0 (C09) & $\sim$0.001$^c$ & 1.77 & 1.30 & 1.01 & 1.55 & 0.77 & 2.88 & 0.54 & 0.78 & 1.33 \\
\hline
   \end{tabular}
$^a$For the models calculated by \citet{bisterzo10} we also need to indicate the choice of the free 
parameter used in those models to describe the efficiency of the $^{13}$C pocket, 
where ST corresponds to the standard case, as described in detail by those authors. This free parameter 
accounts for a possible range of $^{13}$C and $^{14}$N profiles in the $^{13}$C pocket, resulting 
in a range of total integrated neutron fluxes. Thus, it is the main free parameter determining the final
$s$-process abundance distribution in these models.
$^b$Model computed with the opacities by \citet{lederer09}. 
$^c$In the models calculated by \citet{cristallo09a} the $^{13}$C pocket is 
included self-consistently using an overshooting mechanism with the free 
parameter $\beta$=0.1, as described in details in that paper. This results in a mixing 
mass of the order of $\sim$ 0.001 \msun, but decreasing with the TP number, while in our models it 
is 
kept constant. Furthermore, the 
resulting proton profile is not linear in logarithmic scale \citep[see Fig.~4 of][]{cristallo09a}, 
as imposed in our models. 
   \end{sidewaystable}
\end{center}

%% The following command ends your manuscript. LaTeX will ignore any text

\end{document}